%% file: main.tex
\documentclass[10pt, journal, letterpaper]{IEEEtran}
\IEEEoverridecommandlockouts

\usepackage{amsmath,amssymb,amsfonts} % Math packages
\usepackage{graphicx} % For including images
\usepackage{hyperref} % For hyperlinks
\usepackage{cite} % For managing citations
\usepackage{algorithm}
\usepackage{algorithmic}
\usepackage{fancyhdr} % For custom headers/footers
\usepackage[dvipsnames]{xcolor}
\usepackage{subcaption}
\usepackage{comment}
\usepackage{soul}
\usepackage{xurl}
\usepackage{enumitem}

\usepackage{tikz}
\newcommand*\circled[1]{\tikz[baseline=(char.base)]{
            \node[shape=circle,draw,inner sep=2pt] (char) {#1};}}

\renewcommand{\vec}[1]{\boldsymbol{ #1 }}
\newcommand{\Real}{\mathbb{R}}

\usepackage[many]{tcolorbox}
\usepackage{lipsum}

\definecolor{titlebg}{RGB}{100,22,72}
\definecolor{introbg}{RGB}{0,128,128}

\newcommand{\affmark}[1]{\textsuperscript{#1}}
\newcommand{\eqmark}{\textsuperscript{*}}

\newtcolorbox{usecase}[1][]{
  breakable,
  enhanced,
  arc=0pt,
  outer arc=0pt,
  colframe=titlebg,
  colback=titlebg!05,
  overlay unbroken and first={
    \node[
      draw=titlebg,
      fill=titlebg,
      rotate=0,
      anchor=north west,
      text=white,
      font=\bfseries
    ]
    at (frame.north west)  
    {#1};
  }
}

\newtcolorbox{mission}[1][]{
  breakable,
  enhanced,
  arc=0pt,
  outer arc=0pt,
  colframe=introbg,
  colback=introbg!05,
  overlay unbroken and first={
    \node[
      draw=introbg,
      fill=introbg,
      rotate=0,
      anchor=north west,
      text=white,
      font=\bfseries
    ]
    at (frame.north west)  
    {#1};
  }
}

\begin{document}
% Title
\title{Democratizing LLM Training with the Edge: Challenges, Systems, and the Road Ahead}

% Author affiliation and contribution markers
\author{%
Ziyue~Luo\affmark{1},
Jiaxuan~Cai\affmark{1}\eqmark,
Cedric~Le~Denmat\affmark{1}\eqmark,
Srijith~Nair\affmark{1}\eqmark,
Fatemeh~Nourzad\affmark{1}\eqmark,
Rohith~Krishnan~Sudha\affmark{1}\eqmark,
Qinhang~Wu\affmark{1}\eqmark,
Jifan~Zhang\affmark{2}\eqmark,
Sungjae~Lee\affmark{1},
Zhe~Li\affmark{3},
Peiwen~Qiu\affmark{1},
Siddharth~Shah\affmark{4},
Rishabh~Sharma\affmark{2},
Sundararajan~Srinivasan\affmark{5},
Yinglun~Xia\affmark{1},
Xue~Zheng\affmark{1},
Zidong~Liu\affmark{6},
Bicheng~Ying\affmark{7},
Kaushik~Chowdhury\affmark{5},
Gauri~Joshi\affmark{4},
Yingbin~Liang\affmark{1},
Robert~Nowak\affmark{2},
Srinivasan~Parthasarathy\affmark{1},
Saurav~Prakash\affmark{8},
Balaraman~Ravindran\affmark{8},
Sanjay~Shakkottai\affmark{5},
Ness~B.~Shroff\affmark{1},
Haibo~Yang\affmark{3},
Aylin~Yener\affmark{1},
and Jia~Liu\affmark{1}%
\thanks{%
\textsuperscript{1}The Ohio State University;
\textsuperscript{2}University of Wisconsin--Madison;
\textsuperscript{3}Rochester Institute of Technology;
\textsuperscript{4}Carnegie Mellon University;
\textsuperscript{5}The University of Texas at Austin;
\textsuperscript{6}ComboCurve;
\textsuperscript{7}Google;
\textsuperscript{8}Indian Institute of Technology Madras.

\textsuperscript{*}Co-primary authors with equal contribution.
}
\thanks{This work is supported in part by NSF Grant 2112471.}
}

\maketitle

% Abstract
\begin{abstract}
The growing computational power of modern GPU data centers has let a few industrial giants train and deploy large language models (LLMs) on massive amounts of data. 
Yet the sheer scale of these models and their training data makes centralized development increasingly hard to sustain, and at the same time shuts out smaller institutions that hold relevant private data but lack comparable infrastructure. 
In this paper, we organize our positions on distributed LLM training around the pretraining and fine-tuning paradigms, covering federated learning, split learning, and decentralized approaches, and we assess their readiness for geo-distributed settings with heterogeneous hardware and unreliable connectivity. 
We then present a {\em unified} system architecture built around a cloud-hosted parameter server that coordinates heterogeneous clients over outbound gRPC connections, reusing one communication, compression, and fault-tolerance stack for both pretraining and a family of fine-tuning methods. 
We validate the design with real geo-distributed experiments on GPT-2 Medium and Llama3-1B pretraining across multiple university sites, showing stable convergence under realistic conditions. 
Finally, we chart the road ahead, identifying open problems that span frontier model support, elastic fault tolerance, adaptive compression, federated alignment, privacy and trust, incentive design, and system--algorithm co-design, all of which must be addressed to make democratized LLM training a reality.
\end{abstract}

% Keywords

\input{sec/intro-new}
\input{sec/related_work-new}

\input{sec/sys_unified-new}

\input{sec/conclusion-new}

\bibliographystyle{IEEEtran}
\bibliography{
    ./bib/general,
    ./bib/fl,
    ./bib/split_learn,
    ./bib/foundation,
    ./bib/code,
    ./bib/prelim,
    ./bib/distrib_llm,
    ./bib/future_directions
}

\end{document}

%% file: sec/intro-new.tex
\section{Introduction}
\label{sec:intro}
Recent progress in Large Language Models (LLMs)~\cite{Touvron:23:LLaMA,Radford:19:OpenAI,Devlin:19:BERT,OpenAI:24:GPT4} and foundation models~\cite{Radford:21:CLIP,Ramesh:22:DALLE2} has been driven almost entirely by a few large technology companies operating vast, centralized computing centers. This concentration of compute carries three costs. First, it limits \emph{transparency}, since the most capable models are developed behind closed doors, which constrains independent scientific review and replication. Second, it raises \emph{trust and security} concerns, because closed and proprietary training pipelines are difficult to audit. Third, it imposes \emph{environmental} costs, as hyperscale data centers consume enormous amounts of energy and contribute substantially to carbon emissions.

Scaling laws indicate that model quality keeps improving with more data and more parameters~\cite{Kaplan:20:preprint,Hoffmann:22:NIPS}, which only sharpens the incentive to train ever larger models. Yet two bottlenecks increasingly restrict who can participate. The first is \textbf{data availability and governance}: much of the world's useful data is private, and centralizing it raises privacy, bandwidth, and storage costs that many data owners cannot or will not pay. The second is \textbf{computational resources}: a single training run often demands hundreds of high-end GPUs joined by high-throughput interconnects, an infrastructure that lies well beyond the reach of most universities and smaller firms.

Distributed and privacy-preserving paradigms such as federated learning~\cite{mcmahan2017fedavg} and split learning~\cite{Vepakomma:18:preprint} offer a natural starting point for loosening these constraints. Extending LLM \emph{training} to geo-distributed clusters of heterogeneous GPUs, however, exposes a harder problem: real wide-area networks are slow and unreliable, and the institutions that own the GPUs sit behind firewalls, NATs, and incompatible software stacks that no single operator controls. This tension motivates the central question of this paper.

\begin{center}
\begin{tcolorbox}[left=1.pt,right=1.pt,top=1.2pt,bottom=1.2pt,width=1\columnwidth]
\textbf{Key Question:} Can we make LLM training broadly accessible by distributing it across a network, so that training can happen ``anywhere and everywhere''?
\end{tcolorbox}
\end{center}

To answer this question, we observe that LLM development decomposes into two phases, \emph{pretraining} and \emph{fine-tuning}, each with its own objectives and constraints.

\begin{list}{\labelitemi}{\leftmargin=1em \itemindent=-.5em  \itemsep=.2em}

\item {\bf Pretraining across Geo-distributed GPU Clusters:}
Pretraining optimizes a foundation model~\cite{Touvron:23:LLaMA, Radford:19:OpenAI} from scratch, or from a weak initialization, over massive text corpora. We aim to run this phase on underutilized GPUs spread across many institutions. Each site runs a \emph{client} process that trains locally, for example with Fully Sharded Data Parallelism (FSDP)~\cite{zhao2023fsdp}, while coordination across sites takes place over wide-area networks.

\item {\bf Fine-tuning across Geo-distributed GPU Clusters:}
Fine-tuning adapts a foundation model to downstream tasks and adds constraints related to data governance, privacy, and site-specific objectives. Here, each institution fine-tunes on its own data while a central service coordinates the updates, with no need for direct peer-to-peer connectivity between sites.

\end{list}

These two phases differ in their learning objectives, but we argue that a \emph{single system structure} can serve both. That structure is a cloud-hosted parameter server (PS) that holds an authoritative global snapshot, together with a set of heterogeneous clients that perform local updates, periodically push parameter deltas to the PS, and pull refreshed snapshots in return. This abstraction lets us reuse one connectivity model, one fault-tolerance and rejoin path, and one set of communication optimizations for both pretraining and fine-tuning. The two phases differ not in the machinery but in the demands they place on it. \textbf{Pretraining} is dominated by throughput and stability: the system must maximize token throughput under stragglers and WAN delay while maintaining a steady trajectory over very long runs. \textbf{Fine-tuning} is dominated by governance and personalization: updates come from non-i.i.d.\ data, must respect institutional boundaries, and are often parameter-efficient (for example, LoRA) to reduce communication and memory cost.

Despite these differences, both phases run into the same \emph{systems} obstacles once training crosses institutional boundaries over the public Internet. We group them into three challenges that any practical cross-institution training system must confront.

\smallskip
\textbf{Challenge 1: Computation Heterogeneity.}
The first challenge is orchestrating heterogeneous resources, GPUs above all. Even inside a single production cluster, GPU fleets are rarely uniform: new generations arrive quickly, and replacing a cluster wholesale is prohibitively expensive, so operators add newer hardware while keeping older devices in service. The Ohio Supercomputer Center (OSC)~\cite{OSC:25:web}, for example, runs H100s, A100s, and V100s side by side. Using such mixed fleets well already improves training efficiency~\cite{Strati:25:SOSP}, and collaboration across institutions only amplifies the problem, since sites differ in GPU architecture, memory capacity, and software environment. We separate this challenge into hardware and software heterogeneity, then summarize its algorithmic consequences.

\smallskip
\emph{Hardware Heterogeneity.}
Consider one site with a pair of 32~GB V100 GPUs, a second contributing a single 24~GB RTX 4090, and a third offering a bank of A100s with native BF16 support. Running identical work on every machine turns the slower GPUs into \emph{stragglers} that stretch the wall-clock training time and leave faster devices idle. Mixing numerical formats such as FP32, FP16, and BF16 can also produce silent errors when gradients are combined, which in turn risks model divergence.

\smallskip
\emph{Software Heterogeneity.}
Software stacks diverge as well. Driver versions, CUDA toolkits, and NCCL libraries vary across administrative domains; an optimizer built for CUDA~12.2 may fail on CUDA~11.7, and certain vendor BF16 optimizations are simply absent on older cards. Coordinating upgrades across many universities is slow and error-prone, especially when each cluster serves dozens of unrelated research groups.

\smallskip
\emph{Algorithmic Implications.}
Synchronous data-parallel training assumes every step finishes at roughly the same time. Heterogeneous compute breaks that assumption and pushes the system toward asynchronous aggregation. FSDP, designed for homogeneous clusters, does not readily support different shard sizes per device; in practice we instead vary the per-device batch size so that faster GPUs do more work, all while preserving the conditions needed for stable convergence.

\smallskip
\textbf{Challenge 2: Latency and Bandwidth Variability.}
The second challenge is the unpredictable latency and fluctuating bandwidth of a geo-distributed deployment.

\smallskip
\emph{Variability of WAN links.}
Wide-area links between distant institutions and cloud infrastructure are variable and prone to disruption. Packet loss and intermittent connectivity break transfers mid-flight, leaving model synchronization incomplete and training iterations inconsistent. The result is delayed convergence, longer runs, and a less effective system overall.

\smallskip
\emph{Staleness versus Compression.}
Aggressive compression shrinks the bytes on the wire but can introduce bias that slows or destabilizes convergence. Waiting for every worker, on the other hand, lets the slowest link dictate throughput. Prior work on gradient staleness shows that accuracy degrades quickly once delays exceed a few steps for large transformers, yet WAN jitter alone can introduce delays of several seconds. The system must therefore navigate the space between these two failure modes rather than commit to either.

\smallskip
\textbf{Challenge 3: Heterogeneous Network Configurations.}
The third challenge is the diversity of institutional network policies, which we group into three forms.

\smallskip
\emph{Security Middleboxes and Policy Heterogeneity.}
Many institutional clusters sit behind enterprise firewalls or Network Address Translation (NAT) devices that block inbound connections and restrict outbound ports. Edge servers on 5G or LTE change IP addresses often, and campus Wi-Fi may rewrite or cap long-lived TCP connections. Under such varied policies, a simple peer-to-peer overlay is not realistic.

\smallskip
\emph{Unreliable and Asymmetric Paths.}
WAN paths fail for mundane reasons, from fiber cuts to scheduled maintenance, and wireless links fade. Strongly asymmetric bandwidth, where upload and download speeds differ sharply, complicates any protocol that leans on acknowledgments. A reset connection can break a standard gRPC stream and force a costly re-dial that risks dropping in-flight data.

\smallskip
\emph{Operational Complexity.}
Contributors should not have to configure port forwarding or stand up VPN tunnels to participate. Truly \emph{democratized} participation across institutions and edge sites requires a simple ``click-to-join'' experience.

\smallskip
Together, these three challenges define the systems problem of democratized LLM training. Addressing them requires three design principles. First, \emph{heterogeneity awareness}: the system uses asynchronous aggregation that decouples local training from global synchronization, so that stragglers do not stall faster nodes. Second, \emph{asynchronous robustness}: the system bounds staleness and uses fault-tolerant communication, on the assumption that WAN links fail by default rather than as the exception. Third, \emph{network compatibility}: the system relies on a cloud-hosted rendezvous with outbound-only connectivity, which removes the need for peer-to-peer overlays, port forwarding, or VPN setup, and thereby lowers the barrier to entry.

\smallskip
Guided by these principles, this paper makes the following contributions.

\begin{list}{\labelitemi}{\leftmargin=1em \itemindent=-0.5em \itemsep=.2em}
\item We identify and characterize the systems challenges that arise when LLM pretraining and fine-tuning move beyond centralized data centers, namely computation heterogeneity, WAN latency and bandwidth variability, and diverse network configurations, and we argue that these challenges remain unsolved at the scale that democratized training requires.

\item We survey the landscape of distributed LLM training along two axes, training paradigm (pretraining versus fine-tuning) and coordination style (synchronous, asynchronous, and infrequent-synchronization). We examine how existing methods handle compute, network, and statistical heterogeneity, and where they fall short once deployed in geo-distributed, WAN-constrained settings with diverse institutional networks.

\item We propose a unified two-level system architecture in which a cloud-hosted parameter server, talking to heterogeneous clients only over outbound gRPC connections, supplies shared building blocks (periodic delta synchronization, quantization-based compression, and fault-tolerant rejoin) that a federated pretraining paradigm and a federated fine-tuning paradigm reuse without modification. Preliminary results on GPT-2 Medium and Llama3-1B pretraining show stable convergence and competitive wall-clock efficiency under realistic WAN conditions.

\item We describe a set of open problems for the community. These span scaling to frontier and multimodal models, fault tolerance under elastic membership, adaptive communication compression, federated alignment, privacy and trust at LLM scale, incentive design for sustained participation, and system-algorithm co-design for dynamic WAN environments.

\end{list}

To facilitate reproducibility and future research,
we make the implementation of our unified distributed
LLM training framework publicly available at \url{https://github.com/ai-edge-institute/geo-distributed-llm-training}.

The rest of the paper is organized as follows. Section~\ref{sec:related} surveys distributed learning and existing systems for distributed LLM pretraining and fine-tuning. Section~\ref{sec:system} presents our proposed unified architecture, describes the shared infrastructure and the two paradigms it supports, and reports preliminary experiments. Section~\ref{sec:conclusion} discusses open problems and the research agenda ahead. Section~\ref{sec:call-to-arms} concludes the paper with a call to arms.

%% file: sec/related_work-new.tex
\section{Background: Distributed LLM Training}
\label{sec:related}

In this section, we review existing methods for distributed LLM training, grouped into two broad categories: pretraining and fine-tuning.
The underlying distributed optimization techniques, such as federated learning and split learning, are shared across both categories, but the optimization process and the computation and communication requirements differ markedly between them.
The training category also shapes specific system choices, including the use of compression, whether the model is pipelined or split, and the kind of network messages exchanged.

\subsection{Preliminaries}
\label{sec:related_preliminaries}
\input{sec/related/prelim-new.tex}

\subsection{Distributed LLM Pretraining}
\label{sec:related_pretraining}
\input{sec/related/pretraining-new.tex}

\subsection{Distributed LLM Fine-tuning}
\label{sec:related_finetuning}
\input{sec/related/finetuning-new.tex}

%% file: sec/related/prelim-new.tex
\paragraph{Distributed learning vs.\ federated learning}
\label{par:dist_vs_fl}
\emph{Distributed learning} trains a shared model across multiple workers, traditionally in data-center settings with centralized coordination, fast intra-cluster communication, and increasingly sophisticated sharding and parallelism stacks for large models~\cite{dean2012distbelief,li2014ps,rajbhandari2020zero,shoeybi2020megatron,zhao2023fsdp,zheng2022alpa}. \emph{Federated learning (FL)}, by contrast, is defined by organizational boundaries: data remain local to multiple owners, communication is constrained, and training proceeds through exchanged model updates rather than centralized data collection~\cite{konevcny2016federated,mcmahan2017fedavg}. As a result, FL must treat participation, availability, heterogeneity, and system-level issues such as scheduling, fault tolerance, and secure aggregation as first-class concerns~\cite{bonawitz2019system,li2020fedprox}, making the optimization method inseparable from the communication protocol that governs it.

\paragraph{Parameter servers: a foundational abstraction for many-to-one coordination.}
A natural abstraction for large-scale training is the \emph{parameter server} (PS): a logically centralized service maintains model parameters, while workers pull parameters, compute updates, and push them back for aggregation. Early systems such as DistBelief already followed this many-to-one pattern in practice~\cite{dean2012distbelief}, and later work formalized PS as a general framework for distributed learning with flexible consistency models, elastic scaling, and fault tolerance~\cite{li2014ps}. 

PS-style coordination is especially attractive when direct peer-to-peer communication is difficult or undesirable. Because workers communicate primarily with a coordinator rather than a dense all-to-all group, this abstraction fits dynamic membership, heterogeneous workers, and settings with asymmetric or policy-constrained connectivity. It also provides a natural control point for enforcing synchronous, asynchronous, or bounded-staleness update rules~\cite{li2014ps,ho2013ssp}. Although modern data-center training often emphasizes collective communication, systems such as BytePS show that PS-style and collective designs can be viewed within a unified design space~\cite{jiang2020byteps}. For WAN-scale and cross-domain learning, the PS abstraction remains especially appealing because it aligns with restricted connectivity, variable worker availability, and explicit server-side control over synchronization.

\paragraph{Synchronous, asynchronous, and infrequent-synchronization training}
A central axis in both distributed learning and FL is the synchronization model, since it determines both \emph{when} workers communicate and \emph{which version of the model} they optimize against. In \emph{synchronous} training, workers compute in parallel and then coordinate a global update at a barrier, as in data-parallel SGD or one FedAvg communication round~\cite{mcmahan2017fedavg}. This yields clean semantics because all workers update from the same parameter snapshot, but wall-clock progress is limited by the slowest worker and communication path.

In \emph{asynchronous} training, workers instead push updates whenever they finish local computation. The idea predates modern deep learning: early work on asynchronous iterative methods and distributed optimization established convergence under delayed information~\cite{chazan1969chaotic,tsitsiklis1986distributed,bertsekas1989parallel}. Later, \textsc{Hogwild!} showed that lock-free asynchronous stochastic updates can work well in suitable regimes~\cite{recht2011hogwild}, while systems such as DistBelief and the parameter-server line brought asynchronous and partially synchronous execution to large-scale learning~\cite{dean2012distbelief,li2014ps}. The main benefit is higher utilization under stragglers and heterogeneity; the main cost is \emph{staleness}, since an update may be computed from parameters that are already outdated when applied.

This tradeoff motivates \emph{bounded-staleness} models that lie between strict synchronization and unconstrained asynchrony. Stale Synchronous Parallel (SSP) is the canonical example: workers may proceed asynchronously, but only up to a bounded lag, making staleness an explicit system knob~\cite{ho2013ssp}. More recent analyses continue to study asynchronous SGD under heterogeneous computation and communication patterns~\cite{koloskova2022sharper}. In practice, similar ideas also appear through buffered aggregation, time-based synchronization, and stale-update filtering.

The same spectrum appears in FL. Standard FL is usually organized into synchronous \emph{rounds}, where clients train locally and then send updates to a coordinator; FedAvg is the canonical example and already reflects the idea of amortizing communication through multiple local steps~\cite{mcmahan2017fedavg}. Asynchronous FL removes this round barrier and allows clients to contribute whenever they finish local work. Asynchronous Federated Optimization explicitly incorporates stale client updates into FL~\cite{xie2019asyncfl}, while buffered asynchronous aggregation provides a practical middle ground~\cite{nguyen2022fedbuff}.

A closely related design choice is \emph{infrequent synchronization}: instead of communicating often, workers perform more local computation between exchanges. This includes local SGD / periodic averaging, with FedAvg as the FL counterpart~\cite{stich2018localsgd,mcmahan2017fedavg}. The system advantage is lower coordination cost; the optimization risk is greater drift across replicas, especially under non-i.i.d. data, heterogeneous local work, or delayed communication. Recent large-model training systems revisit this idea explicitly in low-communication regimes, with DiLoCo as a representative example~\cite{Douillard:23:DiLoCo}. Overall, synchronous execution, asynchronous execution, bounded staleness, and infrequent synchronization are best viewed as points along a broader \emph{consistency--communication tradeoff}: tighter synchronization reduces staleness but amplifies waiting, while looser or less frequent synchronization improves utilization at the cost of noisier and more drift-prone updates.

\paragraph{Heterogeneity in compute, network, and statistical heterogeneity}
Heterogeneity is a defining feature of federated and WAN-scale learning. It appears at least in three forms: \emph{compute heterogeneity}, where clients differ in hardware, memory, workload, and availability; \emph{network heterogeneity}, where bandwidth, latency, jitter, and reliability vary across links; and \emph{statistical heterogeneity}, where clients hold non-i.i.d. data with different sample sizes or label distributions. These forms are tightly coupled: slower devices and weaker links create stragglers and stale updates, while non-i.i.d. data amplifies the optimization error caused by delayed or uneven participation.

One line of work addresses \emph{systems heterogeneity}. FedProx is a canonical example because it explicitly models FL under both systems and statistical heterogeneity, allowing variable local work while stabilizing training with a proximal term~\cite{li2020fedprox}. TiFL mitigates stragglers through tier-based client grouping~\cite{chai2020tifl}, while Oort improves time-to-accuracy by selecting clients that are both system-efficient and statistically useful~\cite{lai2021oort}. FedScale complements these methods by exposing such effects through realistic large-scale benchmarking~\cite{lai2022fedscale}.

A second line of work targets \emph{statistical heterogeneity}, especially client drift under non-i.i.d. data. FedProx partially addresses this through regularization~\cite{li2020fedprox}; SCAFFOLD uses control variates to correct drift~\cite{Karimireddy:19:CoRR}; FedNova addresses objective inconsistency caused by unequal local work~\cite{wang2020fednova}; and MIME reduces the gap between federated and centralized optimization under heterogeneous data~\cite{karimireddy2021mime}.

For WAN-scale cross-silo training, the key lesson is that these heterogeneities should not be treated in isolation. Systems heterogeneity affects who participates, network heterogeneity affects how delayed updates become, and statistical heterogeneity determines how harmful those delays are after aggregation. The relevant question is therefore not simply how to ``handle heterogeneity,'' but which form of heterogeneity a method addresses, through what mechanism, and under what communication and consistency assumptions.

%% file: sec/related/pretraining-new.tex
Pretraining is a supervised learning method that trains the parameters of an LLM
on large text corpora \cite{Hoffmann:22:NIPS, Groeneveld:24:ACL, Brown:20:NIPS}.
It can be interpreted as a pattern recognition problem, where the model learns
to generate text in a given language, sequentially via \emph{auto-regressive}
generation \cite{Vaswani:17:NIPS}.
Mathematically, suppose the LLM is defined as $\vec{f}_{\vec{\theta}}(\cdot)$,
parametrized by $\vec{\theta}$, that takes as input the length $(k-1)$ sequence
of \emph{tokens} $\{\vec{x}_i\}_{i=1}^{k-1}$ and outputs the next token
$\vec{x}_k$ for some integer $k \geq 2$.
At the next step, $\vec{x}_{k+1}$ is generated by $\vec{x}_{k+1} =
\vec{f}_{\vec{\theta}}(\{\vec{x}_i\}_{i=1}^{k})$, and so on.
Given a large text corpus consisting of $P$ passages of lengths
$\{K_p\}_{p=1}^P$ each $\{\vec{x}_i^p\}_{i=1}^{K_p}$, the pretraining problem
can be formulated as:
\begin{equation}
    \min_{\vec{\theta}} \mathcal{L}(\vec{\theta}) :=
    \frac{1}{P} \sum_{p=1}^P \left[
        \frac{1}{K_p} \sum_{k=2}^{K_p} \mathcal{\ell}_{\mathrm{CE}}(
        \vec{f}_{\vec{\theta}}(\{\vec{x}_i^p\}_{i=1}^{k-1}),
        \vec{x}_k
    ) \right]
    \label{eq:pretrain-loss}
\end{equation}
where $\ell_{\mathrm{CE}}(\cdot, \cdot)$ is the cross-entropy loss function.

\paragraph{Key Challenges with Centralized Pretraining}
While pretraining is a straightforward application of supervised learning
from an optimization perspective, the main challenge with LLM pretraining is
that the training data and the model parameter sizes are unprecedentedly large
when compared to the ML models and datasets used just a decade ago
\cite{Brown:20:NIPS, Hoffmann:22:NIPS, Shazeer:17:ICLR, Liu:24:preprint}.
In fact, the current state-of-the-art (SOTA) LLMs simply cannot be pretrained on
single hardware accelerators (e.g., GPU, TPU or HPU) or a single stand-alone
machine, and instead require several hours (often, days) of distributed
training across powerful data centers with hundreds of high-performance devices.
Besides the optimization process, an often ignored overhead in the centralized
pretraining pipeline, is data collection and curation \cite{Kandpal:25:ICML}.
For these reasons, centralized LLM pretraining has driven a sharp rise in energy expenditure and is affordable only to the largest technology firms.
Almost all SOTA LLMs are therefore proprietary and maintained by a handful of corporations, which leaves LLM pretraining research dominated by industry and public academic work well behind.
Centralized training faces a further limit: scaling laws~\cite{Kaplan:20:preprint} imply that the supply of publicly available data will be insufficient to meet LLM training demands by 2028~\cite{Villalobos:24:ICML}.

\paragraph{Distributed Systems}
These concerns together call for a collaborative effort to build genuinely distributed and decentralized systems for training LLMs at scale.
Towards this end, academics and startups \cite{Gugger:22:Accelerate, Zhu:25:Mist,
Beutel:20:preprint:flower, Sani:24:preprint:photon, Ye:24:openfedllm,
Douillard:23:DiLoCo},
and large corporations \cite{Roth:23:IEEE:flare, zhao2023fsdp,
Douillard:23:DiLoCo} have proposed to leverage existing distributed ML systems
as a means to facilitate a more \emph{democratic approach} to LLM training.
\emph{Federated Learning} (FL) \cite{mcmahan2017fedavg, li2020fedprox,
Karimireddy:19:CoRR, Reddi:20:ICLR} and Decentralized FL (DFL)
\cite{Lalitha:18:NIPS, Savazzi:20:IEEEIoT, Monschein:21:ISCC,
Chellapandi:23:IEEECSL} democratize LLM pretraining on edge devices by forming a
loose federation of participating devices, which in turn train the LLM locally,
\emph{in parallel}, while aggregating models trained by others, either via a
centralized \emph{parameter server}, or in a decentralized manner.
DiLoCo~\cite{Douillard:23:DiLoCo} is the closest to the system we propose and envision here.
Its two defining choices are the use of Nesterov momentum as the outer optimizer and time-based synchronization in place of the step-based synchronization common in conventional FL.
We refer the reader to the surveys~\cite{Lim:20:FLsurvey, Yuan:24:DFLsurvey} for comprehensive coverage of FL and DFL.

There are several key benefits to using such collaborative and decentralized
training paradigms:
\begin{list}{\labelitemi}{\leftmargin=1.7em \itemindent=-0.4em \itemsep=.4em}
    \item From a systems perspective, FL and DFL avoid the cost of collecting
        and centralizing data, since they operate on data held locally by design.
    \item By avoiding data sharing, these methods are intrinsically
        \emph{privacy-preserving}, enabling training on the vast pool
        of private and sensitive data without compromising privacy
        \cite{Wei:20:IEEETIFS}.
    \item Many FL algorithms enjoy \emph{linear speedup} \cite{Yang:21:ICLR},
        meaning that their theoretical convergence rate improves as more
        devices, or clients, participate.
    \item Combined with \ul{S}plit \ul{L}earning (SL)
        \cite{Vepakomma:18:preprint, Thapa:22:AAAI}, smaller clients with much
        lesser compute and memory, can still participate in the training process
        where they train only the initial few layers of the model, leaving the
        bulk of processing to the server.
\end{list}

\paragraph{Open-source Systems}
Several projects such as Ray Anyscale \cite{Moritz:17:Ray}, Amazon SageMaker
\cite{Karakus:21:SageMaker}, Google Vertex AI \cite{Bhatia:23:VertexAI}, NVIDIA
NeMo \cite{Kuchaiev:19:NeMo}, and several others, offer commercial plans or
solutions for distributed LLM training, but are proprietary and run on private
and paid clusters.
Recently, various open-source systems have been designed with support for LLM
pretraining and fine-tuning, such as PrimeIntellect
\cite{Jaghouar:24:PrimeIntellect}.
Flower \cite{Beutel:20:preprint:flower,Sani:24:preprint:photon}, is perhaps the
largest such open-source endeavor, with a large community of contributors, but
many parallel projects have also been initiated in recent years like
\cite{Ye:24:openfedllm, Roth:23:IEEE:flare, Reina:21:preprint:Openfl}.
Some other projects offer decentralized FL systems 
\cite{Cheng:25:preprint:Flock, Singh:26:preprint:Knexa-FL}, and with support for
model pipelining and split learning \cite{Borzunov:23:ACL:petals}.

Most of these systems address a single concern, such as federated or distributed training, or pipelining.
What is increasingly needed, however, is a system that adapts dynamically to system and data heterogeneity and to the geographical layout of the federation.
An ideal system would automatically split or pipeline models that exceed a device's capacity, while performing system-aware and network-aware load balancing and straggler mitigation.

%% file: sec/related/finetuning-new.tex
Fine-tuning is typically the second stage in a two-stage LLM training pipeline,
where the pretrained model is additionally trained, using supervised learning or
reinforcement learning (RL), on a downstream task.
Based on the set of parameters updated, fine-tuning is often classified into 1)
full fine-tuning, where all the LLM parameters are trainable; and, 2) parameter
efficient fine-tuning (PEFT) \cite{hu2022lora, Dettmers:23:NIPS:Qlora,
Li:21:ACL:prefixtune, Ben:21:ACL:BitFit, Pfeiffer:21:ACL:adapterfusion}, where
either a fraction of the parameters or a lower-dimensional representation of the
LLM parameters are trainable while the others remain frozen.
Mathematically, the supervised fine-tuning method can be formulated as
\eqref{eq:pretrain-loss} with the key difference being the reparametrization of
$\vec{\theta}$ using a lower dimensional representation, $\phi$:
\begin{equation}
    \min_{\vec{\phi}} \mathcal{L}\left(\vec{\theta}(\vec{\phi}),
        \{\{\widetilde{\vec{x}}_k^p\}_{k=1}^{K_p}\}_{p=1}^P
    \right),
    \label{eq:finetune-loss}
\end{equation}
where $\mathcal{L}$ is the same objective function as in
\eqref{eq:pretrain-loss}, $\{\{\widetilde{\vec{x}}_k^p\}_{k=1}^{K_p}\}_{p=1}^P$
represents the fine-tuning dataset, $\vec{\phi}$ is typically much smaller in
size than $\vec{\theta}$, and $\vec{\theta}(\vec{\phi})$ is the corresponding parameter set in the original parameter space.
The key characteristic of PEFT is that the gradients computed using
backpropagation are typically of much smaller size, thus requiring lesser
memory.

\ul{L}\ul{o}w \ul{R}ank \ul{A}daptation (LoRA) \cite{hu2022lora} is, by far,
the most heavily studied PEFT method.
Introduced in a centralized learning setup, the approach has also gained a lot
of popularity in FL \cite{wu2026asurvey, Yan:25:IEEEComm}.
The main idea is to replace weight matrices in a model $\vec{W}_i \in \Real^{m
\times n}$ with low-rank updates $\Delta \vec{W}_i$:
\begin{equation}
    \vec{W}_i + \Delta \vec{W}_i = \vec{W}_i + \vec{B}_i \vec{A}_i
    \label{eq:lora-reparam}
\end{equation}
where, the lifting matrices $\vec{B}_i \in \Real^{m \times r}$ and the projection
matrices $\vec{A}_i \in \Real^{r \times n}$ for all layers $i$ form the
trainable set of parameters, $\phi$.
The LoRA rank $r$ is typically much smaller than $m$ and $n$.

\paragraph{FL for LoRA fine-tuning}
Fine-tuning under the FL framework has gained substantial traction lately, largely because training large models on edge or IoT devices with scarce communication and computation budgets is increasingly relevant~\cite{Sun:24:ICLR, Wang:24:NIPS, Bai:24:NIPS:FlexLora}.
Distributed pretraining and fine-tuning share several properties, including local and private data and the absence of a central authority. Fine-tuning in constrained networks, however, calls for additional system-level support such as model splitting~\cite{Vepakomma:18:preprint,Thapa:22:AAAI} or pipeline parallelism~\cite{Harlap:19:sosp:PipeDream, Huang:19:NIPS:GPipe}, since many edge devices cannot store or process an entire LLM and must instead spread it across the network.
Robustness to system, data, and model heterogeneity, along with tolerance for random client dropouts, are further considerations when designing an FL fine-tuning system for such settings.

\paragraph{RL-based fine-tuning}
Fine-tuning methods requiring multi-turn generation are often formulated as reinforcement learning (RL) problems, such as
alignment to human values \cite{Ouyang:22:NIPS:RLHF, Rafailov:23:NIPS:DPO, nourzad2026firm, nourzadbarriers, nourzad2026one}, reasoning for mathematics and coding \cite{Ahn:24:ACL:RLMathsurvey, Jiang:26:ACM:RLCodesurvey}, long-horizon and
tool-calling agentic systems using MCP \cite{Hou:25:ACM:MCP}, retrieval-augmented generation  (RAG) \cite{Gao:23:preprint:RAGsurvey}, etc.
Such systems often require distributed RL across multiple devices \cite{Qi:21:preprint:FedRLsurvey} for the same reasons as pretraining and fine-tuning.

%% file: sec/sys_unified-new.tex
\section{A Unified System for Geo-distributed LLM Training}
\label{sec:system}

In this section, we describe how we take the first step toward realizing the vision of Section~\ref{sec:intro} as a practical, fault-tolerant training system. The system has two levels. The \emph{unified building blocks} (Section~\ref{sec:system-backbone}) provide the shared infrastructure for communication, synchronization, compression, and fault tolerance. The two \emph{application paradigms}, pretraining and fine-tuning (Section~\ref{sec:system-modes}), build on this infrastructure to meet their respective objectives. Throughout the section, we also present preliminary evaluation results that validate the design under realistic geo-distributed conditions.

\subsection{Design Goals}
\label{sec:system-goals}

Our system design is guided by three main goals to address the challenges of distributed LLM training:

\emph{G1: Use heterogeneous GPUs without stragglers.}
The system must keep all participating clusters actively working at the same time, even if they have different computing power (FLOPS), memory capacity, or support for mixed-precision calculations. This goal directly addresses Challenge~1 (computation heterogeneity).

\emph{G2: Handle WAN latency and bandwidth variations.}
Model synchronization cannot assume fast, sub-millisecond network connections. Instead, it must function effectively with links where speeds can change by a large margin. This goal directly addresses Challenge~2 (latency and bandwidth variability).

\emph{G3: Working with firewall/NAT asymmetry and preemptions.}
Clients should only need to make outbound connections. New or re-launched nodes must be able to rejoin the training quickly with a single remote procedure call (RPC) to the parameter server (PS). This goal directly addresses Challenge~3 (diverse network configurations).

\subsection{Level~1: Unified System Building Blocks}
\label{sec:system-backbone}

We first describe the shared infrastructure that is common to both pretraining and fine-tuning paradigms. This comprises the architecture overview, the communication protocol, the network compression scheme, and the fault-tolerance mechanism.

\subsubsection{Architecture Overview}
\label{sec:arch}

\begin{figure}[t]
  \centering
  \includegraphics[width=\linewidth]{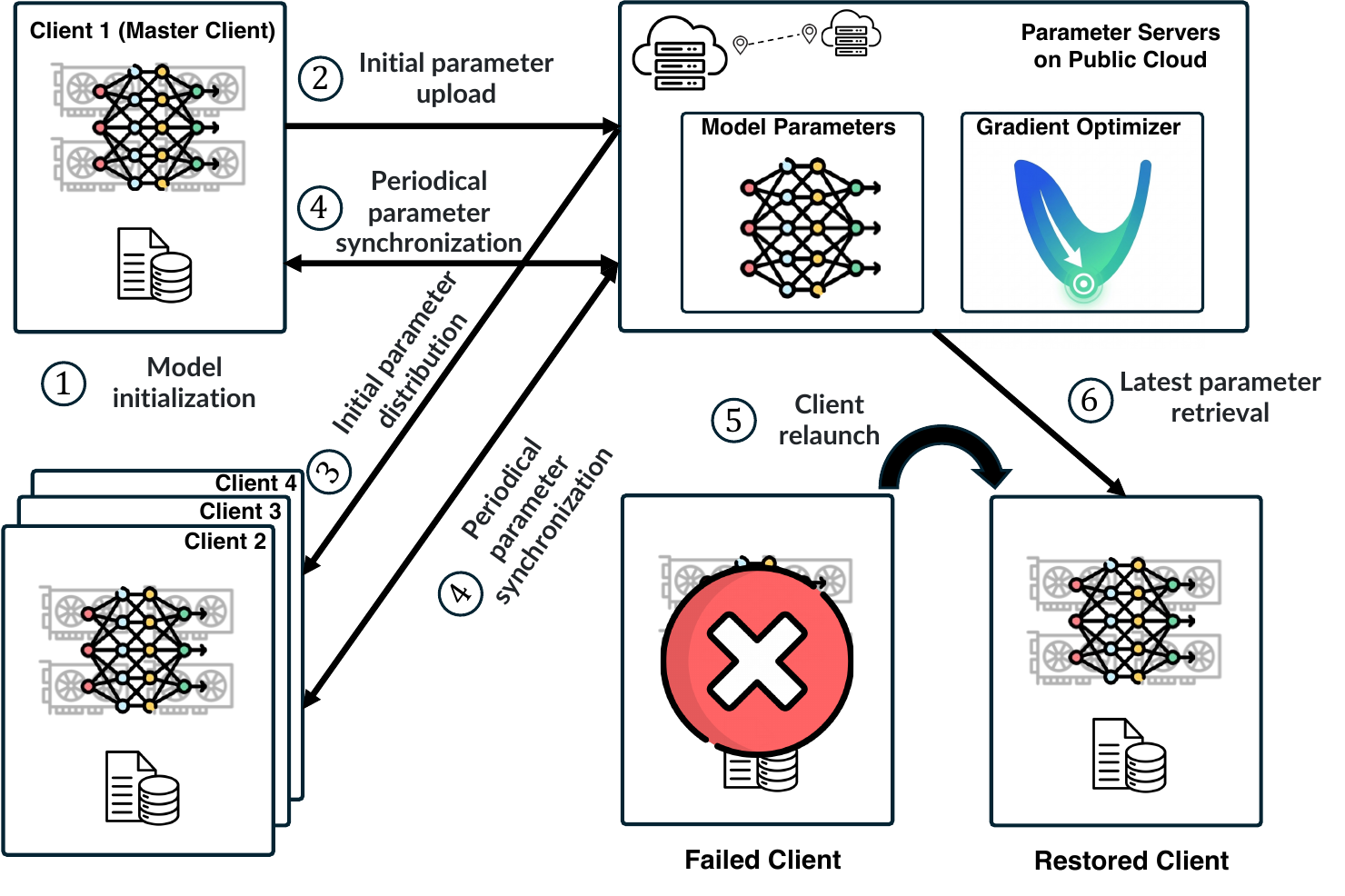}
  \caption{Cross-institution unified training system. A master client initializes the model and uploads it to a cloud parameter server (PS), which maintains the global model and optimizer state. Worker clients periodically synchronize parameters with the PS during training and can recover the latest model after failures. Circled numbers correspond to the protocol steps described in Section~\ref{sec:system-protocol}.}
  \label{fig:system}
\end{figure}

Figure~\ref{fig:system} illustrates our cross-institution unified training system. A cloud-hosted parameter server (PS) holds the authoritative global model parameters and a gradient optimizer. A designated master client (Client~1) bootstraps the model, and any number of additional clients located at campus or edge sites perform local model updates.

\begin{figure}[t]
  \centering
  \includegraphics[width=\linewidth]{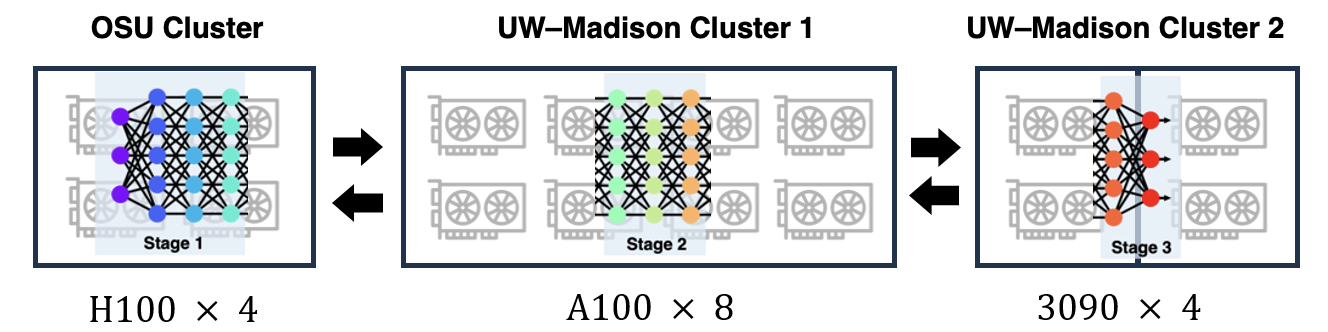}
  \caption{Training with pipeline parallelism. Each stage is trained using data parallelism with AllReduce within each cluster.}
  \label{fig:pipline}
\end{figure}

\paragraph{Design evolution: from pipeline parallelism to PS}
Our initial architectural vision involved ``pipeline parallelism,'' where different levels of the LLM would reside on separate clusters as depicted in Figure~\ref{fig:pipline}. This design was attractive because it allowed scaling very large models beyond the memory capacity of a single cluster and promised reduced communication overhead compared to full data parallelism. However, this pipeline design inherently necessitates direct peer-to-peer (P2P) connections between participating clusters. In practice, the firewalls and network address translation (NAT) policies commonly enforced in institutional networks make direct P2P difficult to implement reliably. This led us to evolve our architecture toward the centralized PS model in Figure~\ref{fig:system}, which provides a more robust and manageable communication infrastructure.

\paragraph{Parameter server (PS)}
We set up our PS as a lightweight gRPC service on Amazon Web Services (AWS). This PS receives parameter differences (deltas) from the clients, applies federated averaging (FedAvg)~\cite{mcmahan2017fedavg} to combine these updates, and sends out new global model snapshots. All communication, including control messages and model tensors, uses outbound connections from clients to the PS, avoiding the need for complex inbound firewall configurations at client sites. The PS aggregates tensor updates in its memory, efficiently streaming data to minimize computing and memory overhead.

\emph{Why a centralized PS?}
Beyond sidestepping the P2P difficulties above, the centralized relay brings two concrete benefits. It reduces the connectivity structure from \(O(N^2)\) peer links to \(O(N)\) client-to-server links, and it provides a single place to manage TLS encryption and access control for the whole federation.

\paragraph{Multi-PS distributed learning} \label{par:multi-ps}
A single central PS can itself become a coordination bottleneck. We therefore offer a multi-PS mode as a configurable option: clients form clusters, typically by geographical proximity, and each cluster is coordinated by its own PS. Training within a cluster proceeds as in Algorithm~\ref{alg:meta-alg}, with the local PS sharing the global model, the clients training on their local data, and the PS aggregating periodically. The PSs then synchronize with one another via ring all-reduce. This setup balances load and mitigates the latency imposed by the geographical layout of the federation, directly addressing goal G2 and partly addressing goal G1.

\paragraph{Client}
Each client starts an FSDP~\cite{zhao2023fsdp} worker. The size of its micro-batches and its mixed-precision mode are automatically adjusted to match the client's local memory and computing power (FLOPS). Clients store their optimizer states locally. Only the \emph{differences} (deltas) from the last global model snapshot are sent over the WAN.

\paragraph{Synchronization cadence}
Instead of synchronizing after every training step, each client sends updates after a configurable wall-clock interval (e.g., 10 minutes). This limits how ``stale'' the model updates can become while also controlling the load on the WAN.

\subsubsection{Communication Protocol}
\label{sec:system-protocol_impl}

Our initial implementation used PyTorch RPC for cross-institution communication. While PyTorch RPC is highly efficient within a single, tightly controlled cluster, its design mandates a mesh-like topology for distributed processes. This typically requires dynamic allocation of numerous ports, potentially leading to $O(N^2)$ connections for $N$ clients. This design conflicted directly with the strict firewall policies common in institutional networks, which often block dynamic port ranges or require explicit, static rules for inbound connections.

In response, we transitioned to the gRPC protocol. gRPC allowed us to explicitly manage ports and precisely define packet formats using Protocol Buffers. A key advantage of gRPC is its ability to operate over a single, fixed, user-specified port via persistent, outbound client connections, thereby circumventing inbound firewall restrictions at client sites. We leveraged gRPC's bidirectional streaming RPCs (e.g., for \textsc{StreamModel} and \textsc{PushDelta} calls) to enable continuous exchange of model updates without constantly re-establishing connections. This strategic change effectively addressed the practical and technical constraints we faced, enabling reliable and secure communication for our distributed training framework.

\subsubsection{Network Compression}
\label{sec:compression}

To reduce network traffic and latency, we compress the gradient payload (addressing Challenge~2). Compression proceeds in two stages: quantization followed by Huffman encoding. Quantization maps each 32-bit floating-point gradient to an 8-bit or 16-bit integer, for compression factors of 4 and 2, respectively. We use min-max quantization for its low overhead, since it requires storing only the minimum and maximum of the gradient distribution. Huffman encoding then compresses the quantized gradients further. The receiver reverses the pipeline, decoding and then dequantizing to reconstruct the payload.

We observed that maintaining min-max quantization pairs for each multi-dimensional tensor resulted in high storage overhead. Our solution is to flatten all multi-dimensional tensors into a single one-dimensional tensor and then perform quantization. Treating each tensor separately for Huffman encoding would not give enough gradient values to obtain a skewed distribution, and flattening the quantized tensors solved this issue.

\subsubsection{Fault Tolerance and Rejoin Logic}
\label{sec:ft}

Interruptions, such as a dropped gRPC stream, signal a preemption. When a client relaunches, it authenticates with a one-time token, then fetches the latest global model ($\Theta$) and its last saved optimizer state from the PS. Because this path retrieves the live model directly, a client typically resumes within roughly 10 seconds. The mechanism preserves convergence guarantees while hiding the failure from the rest of the federation.

For durability beyond a single relaunch, we also checkpoint model and optimizer state to disk as a single file. Checkpointing operates at two levels: a checkpoint is written after every synchronization to guarantee fault tolerance, and a checkpoint manager separately retains the ten checkpoints with the lowest evaluation loss, refreshed every 1000 steps. Recovering from a cold disk checkpoint, rather than from the live PS model, takes about one minute. All checkpoints are stored locally on the client. If the PS itself fails, we relaunch the federation and Client~1 loads its saved checkpoint and streams it to the PS to resume training.

\subsubsection{Protocol Walk-through}
\label{sec:system-protocol}

We illustrate the step-by-step flow of our system using the numbered arrows in Figure~\ref{fig:system}.

\begin{itemize} 
    \item \textbf{\circled{1} Model initialization.} A designated master client (Client~1) initializes the starting model ($\Theta_0$) from a pretrained checkpoint or random weights and prepares its local FSDP shards.

    \item \textbf{\circled{2} Initial parameter upload.} Client~1 streams $\Theta_0$ to the PS using a single \textsc{PushDelta} call. The PS stores $\Theta_0$ in host memory alongside its gradient optimizer state.
    
    \item \textbf{\circled{3} Initial parameter distribution.}
    The PS distributes $\Theta_0$ to all other clients through their open \textsc{StreamModel} pipes. This ensures every client starts from an identical model state.
    
    \item \textbf{\circled{4} Periodic parameter synchronization.} Each client trains on its local data, then after the configured wall-clock interval,
    \emph{(i)} computes the difference ($\Delta$) between its local model ($\theta$) and the global model ($\Theta$),
    \emph{(ii)} quantizes $\Delta$ based on the observed network link rate, and
    \emph{(iii)} sends the quantized $\Delta$ to the PS.
    The PS de-quantizes the received delta, applies FedAvg, and streams the updated global model ($\Theta$) back to the client. The local runtime adjusts the micro-batch size and mixed-precision settings to fully utilize each client's GPU without risking out-of-memory errors.

    \item \textbf{\circled{5} Client relaunch.} If a client is preempted or crashes, it relaunches and re-authenticates with the PS using a one-time token.

    \item \textbf{\circled{6} Latest parameter retrieval.} The restored client fetches the latest global model ($\Theta$) from the PS, loads its last saved optimizer states, and resumes training as described in Section~\ref{sec:ft}.

\end{itemize}

\subsection{Level~2: Application Paradigms}
\label{sec:system-modes}

\begin{figure}[t]
  \centering
  \includegraphics[width=\linewidth]{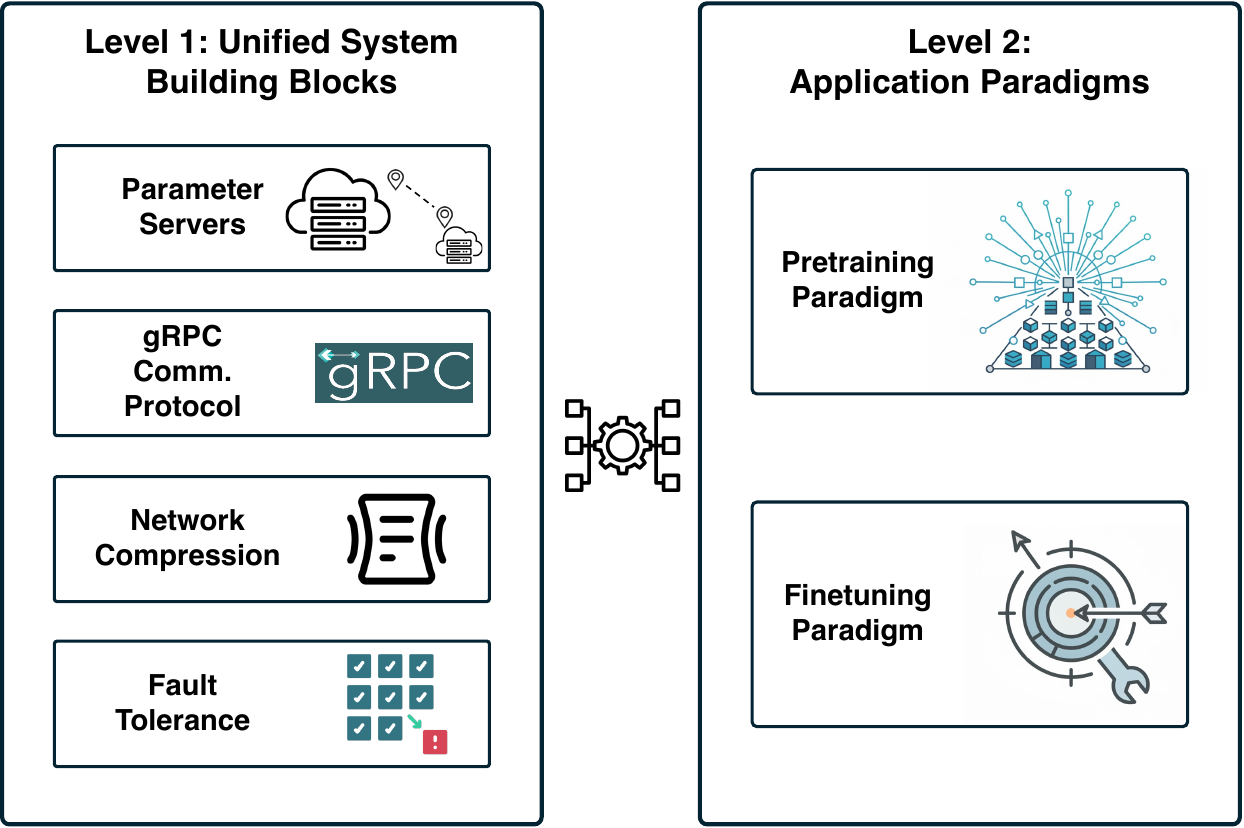}
  \caption{Two-level architecture of our unified system. The unified building blocks (Level~1) provide shared infrastructure for communication, synchronization, compression, and fault tolerance. Both pretraining and fine-tuning (Level~2) build on the same infrastructure, configuring it for their respective objectives.}
  \label{fig:2layer_system}
\end{figure}

The unified building blocks in Figure~\ref{fig:2layer_system} support two learning paradigms that share one infrastructure but differ in objective. This section explains how each paradigm configures that infrastructure, presents a high-level algorithm, and discusses paradigm-specific considerations. We first give a system-level meta-algorithm that serves both pretraining and fine-tuning, then describe the two paradigm implementations in turn.

\subsubsection{A meta-algorithm for distributed LLM training}
Algorithm~\ref{alg:meta-alg} states a general meta-algorithm for training LLMs in a distributed manner. The same underlying system drives distributed pretraining as well as fine-tuning with LoRA (FedIT~\cite{Zhang:24:ICASSP:FedIT}), Hybrid LoRA, RAVAN~\cite{Raje:25:NIPS:RAVAN}, and zeroth-order optimization~\cite{Li:25:ICLR:ZeroOrder}. The meta-algorithm follows conventional FedAvg~\cite{mcmahan2017fedavg}: the parameter server (PS) initializes the global LLM and reparametrizes it into a combination of frozen and trainable parameters.

In LoRA~\cite{hu2022lora}, for instance, each weight matrix is reparametrized as in \eqref{eq:lora-reparam}. The same formulation covers pretraining trivially by setting $\vec{\phi}(\cdot) = \vec{\theta}(\cdot) = \vec{I}(\cdot)$, the identity operator. The PS distributes only the trainable parameters to the clients. Each client freezes the underlying LLM, trains the trainable parameters on mini-batches sampled from its local dataset, and returns the trainable deltas $\vec{\delta}^c$ to the PS. The PS aggregates these deltas and updates the global model with a global learning rate. This process repeats until a convergence criterion is met.

\input{sec/algs/meta-algorithm-new.tex}

\subsubsection{\textbf{Paradigm~1}: Federated Pretraining}

\label{sec:system-pretrain}

\begin{figure}[ht]
    \centering
    \begin{subfigure}{0.39\columnwidth}
        \centering
        \includegraphics[width=\textwidth]{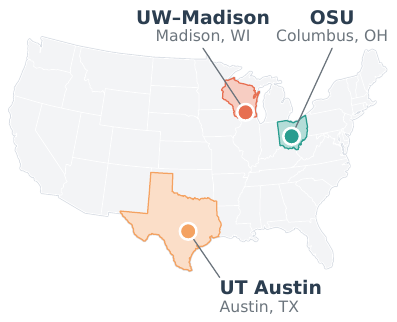}
        \caption{Geo-locations}
    \end{subfigure}
    \begin{subfigure}{0.59\columnwidth}
        \centering
        \includegraphics[width=\textwidth]{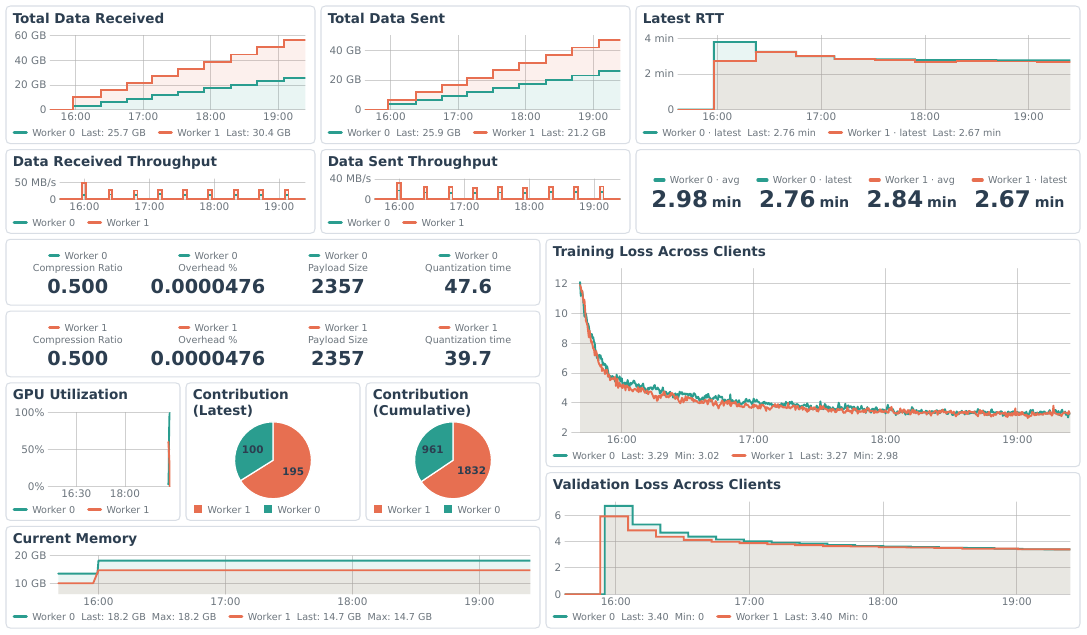}
        \caption{Grafana-based Dashboard}
    \end{subfigure}
    \caption{\label{fig:pretrain-dashboard}A real geo-distributed deployment of our system, \emph{pretraining} GPT-2 Medium across three nodes. Panel~(a) shows the physical locations of the GPU servers across the United States, and panel~(b) shows the live training metrics tracked during the run.}
\end{figure}

Our pretraining system is directly derived from the meta-algorithm in Algorithm~\ref{alg:meta-alg}, by replacing the reparametrization in lines \ref{alg-line:theta2phi} and \ref{alg-line:phi2theta} with the identity transformation, i.e., all parameters are communicated and trained.
Within each client, we leverage gradient accumulation and FSDP~\cite{zhao2023fsdp} to distribute training across multiple GPUs, allowing each worker to process multiple micro-batches and aggregate them into a single mini-batch before performing a parameter update.
This allows each client to select a micro-batch size that fits its hardware constraints, avoiding out-of-memory errors while maintaining a consistent effective mini-batch size across heterogeneous clients (addressing Challenge~1).
Before transmission, gradients are quantized to BF16 or INT8 depending on the observed network link speed (addressing Challenge~2).
If a failure occurs during communication, the system detects the error and initiates immediate retransmission (addressing Challenge~3).
Additionally, to prevent stragglers, i.e., weaker clients with low compute speeds, from slowing down the overall process, we also introduce a timeout $T$ at the PS.
If some clients time out before performing $H$ optimization steps locally, they share their current set of parameters to the PS.
Figure~\ref{fig:pretrain-dashboard} shows the geographical distribution of the GPU servers used for pretraining alongside a real-time Grafana dashboard that tracks training progress together with network latency and compression statistics.

\paragraph{Preliminary results on GPT-2 Medium pretraining}
\label{sec:eval:gpt2}
We evaluate convergence behavior when pretraining GPT-2 Medium~\cite{Radford:19:OpenAI} on OpenWebText using two university sites as workers: OSU (one worker, 2$\times$ NVIDIA H100) and UW--Madison (two workers, each with 2$\times$ NVIDIA RTX 3090), coordinated through an AWS EC2 t2.xlarge instance. Model updates are synchronized every 10 minutes.

We compare with three representative baselines:
(1)~\textbf{FedAvg}~\cite{mcmahan2017fedavg};
(2)~\textbf{DiLoCo}~\cite{Douillard:23:DiLoCo}, which reduces synchronization frequency while leveraging Nesterov momentum~\cite{Sutskever:13:ICML};
(3)~\textbf{FedNova}~\cite{Wang:20:FedNova}, which normalizes local updates to account for heterogeneous worker speeds.
FedAvg and DiLoCo follow a synchronous training paradigm, where all clients are required to complete the same number of local steps before aggregation, making the overall process vulnerable to stragglers, whereas FedNova allows asynchronous updates across clients, similar to our design.

\begin{figure*}[!t]
    \centering
    \begin{minipage}[t]{0.3\textwidth}
        \includegraphics[width=\textwidth]{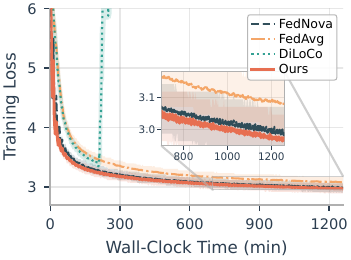
        }
        \caption{GPT-2 Medium training loss.}
        \label{fig:eval:gpt2_loss}
    \end{minipage}
    \begin{minipage}[t]{0.33\textwidth}
        \includegraphics[width=\textwidth]{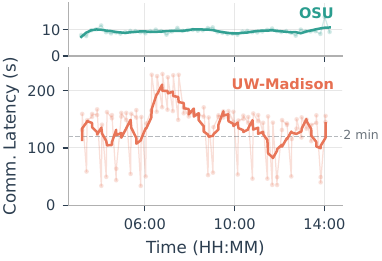}
        \caption{Gradient receive latency across geo-distributed sites.}
        \label{fig:eval:osu_wisc_latency}
    \end{minipage}
    \begin{minipage}[t]{0.3\textwidth}
        \includegraphics[width=\textwidth]{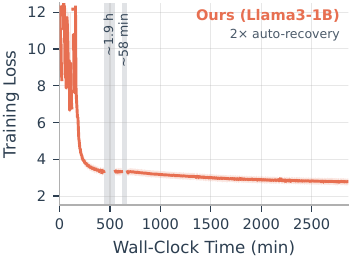}
        \caption{Llama3-1B pretraining loss.}
        \label{fig:eval:llama_loss}
    \end{minipage}
\end{figure*}

Figure~\ref{fig:eval:gpt2_loss} shows the training loss trajectories. Our design converges stably and reduces the loss faster in wall-clock time than the baselines. DiLoCo, in contrast, diverges quickly under its recommended hyperparameters. This illustrates a more general point: {\em algorithms tuned for high-bandwidth, homogeneous data-center environments often perform poorly in geo-distributed deployments, where the latency variance and compute heterogeneity are much higher.} These results indicate that a design must account explicitly for heterogeneous, geo-distributed conditions.

\paragraph{Preliminary results on cross-site communication latency}
\label{sec:eval:latency}
We characterize cross-site communication latency to quantify the practical challenges of geo-distributed training. Figure~\ref{fig:eval:osu_wisc_latency} plots the gradient receive latency measured between our AWS instances and the university sites. We observe two patterns. First, the network conditions fluctuate substantially over time, due to transient congestion and routing changes. Second, latency and bandwidth differ across sites, reflecting differences in local network infrastructure and institutional policies. Together, these effects produce both high average latency and high variance, which create stragglers and destabilize synchronization. They also motivate two of our design choices: a centralized relay that simplifies connectivity under firewall constraints, and communication compression that reduces the payload size and thus the WAN overhead.

\paragraph{Preliminary results on Llama3-1B pretraining}
\label{sec:eval:llama}
We evaluate our system on Llama3-1B~\cite{Touvron:23:LLaMA} to demonstrate that the design scales to more demanding workloads. We use OSU (one worker, 2$\times$ H100) and UT Austin (one worker, 4$\times$ H100), coordinated through two AWS EC2 r5.2xlarge instances, using our multi-PS setup in Section \ref{par:multi-ps}. Model updates are synchronized every 20 minutes with gradient quantization from FP32 to UINT8.

Figure~\ref{fig:eval:llama_loss} plots the training loss trajectory. Despite the larger model and the heavier communication load, the loss decreases steadily and without divergence. This shows that the system remains stable for larger models under realistic WAN conditions.

\subsubsection{\textbf{Paradigm~2}: Federated Fine-tuning}
\label{sec:finetune}

\begin{figure}[ht]
    \centering
    \begin{subfigure}{0.39\columnwidth}
        \centering
        \includegraphics[width=\textwidth]{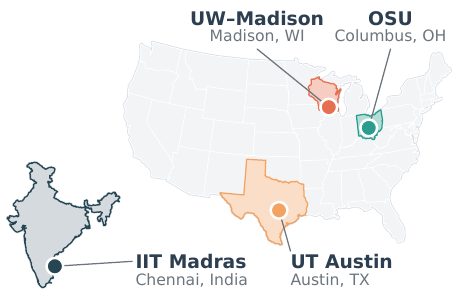}
        \caption{Geo-locations}
    \end{subfigure}
    \begin{subfigure}{0.59\columnwidth}
        \centering
        \includegraphics[width=\textwidth]{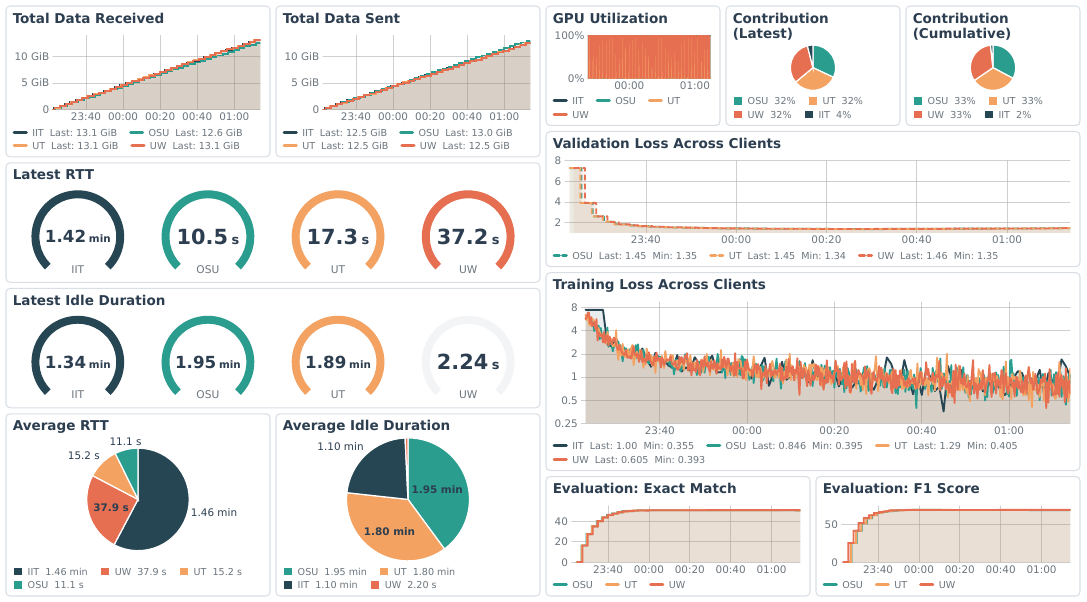}
        \caption{Grafana-based Dashboard}
    \end{subfigure}
    \caption{\label{fig:finetune-dashboard}A real geo-distributed deployment of our system, \emph{Hybrid LoRA} fine-tuning of GPT-2 Medium across three nodes. Panels are read as in Figure~\ref{fig:pretrain-dashboard}.}
\end{figure}

We implement and evaluate five distributed fine-tuning algorithms, each offering different trade-offs in communication efficiency, computation efficiency and privacy.
At the system level, the first four methods use the same system backbone as in Algorithm~\ref{alg:meta-alg}, with different reparametrizations in steps \ref{alg-line:theta2phi} and \ref{alg-line:phi2theta}.
The last method is a Federated Split Learning (FSL) algorithm, which implements specific system changes and diverges from the basic framework introduced in Algorithm~\ref{alg:meta-alg}.
For performance evaluation, we fine-tune the GPT2-medium model for the finetuning task of \emph{question answering}, using the Stanford Question Answering Dataset (SQuAD)~\cite{rajpurkar-etal-2016-squad}.
As an example, Figure~\ref{fig:finetune-dashboard} shows the geographical locations of the GPUs used for LoRA fine-tuning, together with a real-time Grafana dashboard tracking training, validation, and network metrics.

\paragraph{Federated LoRA-based fine-tuning}
In this algorithm, we implement a version of FedIT \cite{Zhang:24:ICASSP:FedIT}.
Instead of updating the full model, we update and exchange only the low-rank adapter matrices $\vec{A}$ and $\vec{B}$ associated with LoRA \cite{hu2022lora} (c.f. \eqref{eq:lora-reparam}).
This method can be realized by using $\vec{\phi} = \{(\vec{A}_l, \vec{B}_l)\}_{l=1}^L$ for $L$ layers in the model, as the reparametrization in Step~\ref{alg-line:theta2phi} in Algorithm~\ref{alg:meta-alg}.
In Algorithm~\ref{alg:meta-alg}, at each federated round, the server aggregates and distributes only $\vec{A}$ and $\vec{B}$ in Step \ref{alg-line:theta2phi}.
Note that the naive aggregation step in FedIT is known to be inconsistent.
However, our system is easily extendable to other variants of FL fine-tuning, such as \cite{Wang:24:NIPS}.

To strengthen privacy in federated LoRA fine-tuning, we also implement \emph{Hybrid LoRA}, which separates global from local updates. The shared reparametrization $\vec{\phi}$ is the identity, exactly as in pretraining, so that the base parameters $\vec{\theta}$ are the only quantities exchanged with the PS. Each client additionally maintains \emph{private} LoRA adapters $\vec{A}$ and $\vec{B}$ that never leave the client. During a round, the client updates the base parameters for $H_{\texttt{base}}$ steps and the private adapters for $H_{\texttt{lora}}$ steps, then returns only the base parameters for aggregation. The private adapters let each site personalize without exposing its adaptation to the federation.

\paragraph{Multi-headed LoRA Adapters (RAVAN)}
We can extend our system backbone to also implement a variant of LoRA known as RAVAN \cite{Raje:25:NIPS:RAVAN}.
In RAVAN, the weights are reparametrized by:
\vspace{-0.7em}
\begin{equation}
\vec{W} + \Delta \vec{W} = \vec{W} + \vec{A} \left(\sum_{i=1}^{h}\vec{H}_i\right) \vec{B},
\end{equation}
where $\vec{H}_i$ are trainable \emph{heads}, and $\vec{W}, \vec{A}$, and $\vec{B}$ are frozen and only shared once in the beginning.
The reparametrization $\vec{\phi} = \{\vec{H}_i\}_i$ extends Algorithm~\ref{alg:meta-alg} to RAVAN.

\paragraph{Zeroth-order federated fine-tuning}

Zeroth-order optimization eliminates the need to communicate gradients back and forth from the server, by using random directional gradients to approximate true gradients.
Consider the reparametrization $\vec{\phi} = (s, g)$, where $s$, a \ul{r}andom \ul{n}umber \ul{g}enerator (RNG) seed, and $g$ a scalar directional gradient.
The idea is to simply transmit the RNG seed from the PS to clients.
The clients then update the parameters using a directional derivative $g_c$ along a random direction, obtained using the shared seed $s$.
The clients share $g_c$ with the server which aggregates it, and then applies the update along the chosen random direction. 
This drastically minimizes the overall communication costs from $\mathcal{O}(d)$ for $d$ parameters, to $\mathcal{O}(1)$.
We refer the interested reader to \cite{Li:25:ICLR:ZeroOrder} for details on zeroth-order FL fine-tuning.

\paragraph{Federated split learning (FSL-SAGE)}
\label{sec:fsl_sage_mode}

Conventional FL assumes that every client can store and process the full model, an assumption that breaks down for very large models on resource-constrained clients. Federated Split Learning (FSL)~\cite{Vepakomma:18:preprint, Thapa:22:AAAI} addresses this by partitioning the model between the client and the server. To cut both the communication load and the number of synchronization cycles, we adopt FSL-SAGE~\cite{Nair:25:ICML}, a provably convergent, communication-efficient FSL algorithm built around auxiliary models. FSL-SAGE uses two server processes: an $F$-server orchestrates federated aggregation of the client-side models, and an $S$-server trains the server-side model and periodically realigns each client's auxiliary model. Because auxiliary-model alignment runs far less often than the per-iteration synchronization that vanilla SL requires, FSL-SAGE substantially lowers communication cost and lets the clients run in parallel.
We deploy the $F$-server and $S$-server on separate AWS instances, with clients on single-GPU servers at The Ohio State University (OSU), University of Michigan (UMich), The University of Texas at Austin (UT Austin), and Indian Institute of Technology Madras (IIT Madras), and apply the system to LoRA fine-tuning of GPT-2 Medium for question answering on SQuAD~\cite{rajpurkar-etal-2016-squad}. To our knowledge, this is the first geo-distributed framework for scalable federated split learning.

%% file: sec/algs/meta-algorithm-new.tex
\newcommand{\client}{c}
\newcommand{\ps}{\texttt{PS}}

\begin{algorithm}[t]
\caption{\label{alg:meta-alg} General meta-algorithm for Distributed Training of LLMs}
\begin{algorithmic}[1]
    \STATE \textbf{Input:} LLM with trainable params $\vec{\theta}_0$, local steps $H$, sync timeout $T$, global learning rate $\eta$
    \STATE Parameter server (PS) initializes $\vec{\theta}^{\ps} \gets \vec{\theta}_0$
    \STATE \label{alg-line:theta2phi} \textcolor{ForestGreen}{Reparametrize and extract trainable parameters from $\vec{\theta}^{\ps}$: $\vec{\phi}^{\ps} := \vec{\phi}(\vec{\theta}^{\ps})$}
    \REPEAT
        \STATE PS distributes $\vec{\phi}^{\ps}$ to clients.
        \FOR{each client $\client$ \textbf{in parallel}}
            \STATE Initialize $\vec{\phi}^{\client} \gets \vec{\phi}^{\ps}$
            \STATE Locally update $\vec{\phi}^{\client}$ until $H$ steps or timeout $T$, whichever comes first
            \STATE Compute and send deltas $\vec{\delta}^{\client} := \vec{\phi}^{\client} - \vec{\phi}^{\ps}$ to PS
            \STATE \textbf{If interrupted: } Request $\vec{\phi}^{\ps}$ from PS and restart local training
        \ENDFOR
        \STATE PS receives and aggregates all client models $\vec{\phi}^{\ps} = \vec{\phi}^{\ps} + \eta . \frac{1}{N} \sum_{i=1}^N \vec{\delta}^{\client}$
        \STATE \label{alg-line:phi2theta} \textcolor{ForestGreen}{Restore and checkpoint global model $\vec{\theta}^{\ps} = \vec{\theta}(\vec{\phi}^{\ps})$}
    \UNTIL{terminated}
    \STATE \textbf{Output:} Global model $\vec{\theta}^{\ps}$
\end{algorithmic}
\end{algorithm}

%% file: sec/conclusion-new.tex
\section{The Road Ahead}
\label{sec:conclusion}

In the previous sections, we described a unified system that runs both pretraining and fine-tuning across geo-distributed, heterogeneous GPU clusters, and we showed that it converges stably under realistic WAN conditions. The system is, however, only a first step. Several questions remain before such an approach can train frontier-scale models, absorb the churn of an open federation, and reward participants in a way that sustains their contribution over long runs. In this section, we describe these open problems. We organize them from the models one would like to train, to the systems and algorithms needed to support them, and finally to the economic structures that determine whether a federation can be sustained at all.

\subsection{Scaling to Frontier and Multimodal Models}
Our experiments consider models up to 1B parameters. Scaling to frontier sizes, on the order of 70B parameters and beyond, will require new strategies rather than larger versions of the present ones.
One promising direction is \emph{Mixture-of-Experts} (MoE) architectures~\cite{Shazeer:17:ICLR, Liu:24:preprint}, whose sparse activation reduces the per-client compute and memory footprint. In a geo-distributed setting, however, the placement and routing of experts across WAN-connected sites becomes a difficult systems problem in its own right, since a naive cross-site partition can inflate communication, unbalance load, and destabilize routing.
A second direction is to extend the system to \emph{multimodal} foundation models such as LLaVA~\cite{Liu:23:NeurIPS:LLaVA}. Such models appear compatible with our framework, but they introduce additional heterogeneity in their modality-specific encoders, in data availability, and in communication cost that the current design does not yet address.

\subsection{Fault Tolerance under Elastic Membership}
Fault tolerance in large-scale training has received considerable attention. Bamboo~\cite{Thorpe:23:NSDI:Bamboo} tolerates frequent spot-instance preemptions by filling pipeline bubbles with redundant layer computation; Oobleck~\cite{Jang:23:SOSP:Oobleck} pre-computes a set of heterogeneous pipeline templates so that survivors can quickly recover a working configuration; and Gemini~\cite{Wang:23:SOSP:Gemini} reduces recovery latency by more than 13$\times$ using multi-tiered in-memory checkpointing. These systems assume managed clusters with reliable, high-bandwidth interconnects and homogeneous hardware. These assumptions do not hold at the open edge.

In our setting, disruption is the common case rather than the exception. Clients leave and rejoin between rounds, WAN links fluctuate without warning, and new participants arrive during training with unfamiliar hardware. Supporting such \emph{elastic membership} requires reassigning work as clients come and go, moving model state over high-latency links, and bringing new joiners up to the current state without stalling the rest of the federation. We believe that lightweight, WAN-aware failure detection and state reconciliation, designed for open and heterogeneous membership rather than for closed clusters, are the main pieces still missing.

\subsection{Adaptive Communication Compression}
Our system already compresses traffic using quantization and Huffman coding (Section~\ref{sec:compression}). Communication will only become more important as models grow. Error-feedback mechanisms~\cite{Karimireddy:19:CoRR} and adaptive quantization~\cite{RamezaniKebrya:21:JMLR:NUQSGD} can reduce traffic further. Interestingly, the ``Square-Cube Law'' observed in SWARM parallelism~\cite{Ryabinin:23:ICML:SWARM} shows that in some regimes larger models are in fact \emph{more} communication-efficient, because computation grows faster than communication volume.
The open question is how to select the compression strategy \emph{adaptively}, choosing bit-widths, sparsification ratios, and synchronization frequencies from real-time network measurements. We expect that doing so will require \emph{co-designing} the compression scheme with the outer optimizer~\cite{Douillard:23:DiLoCo}, so that the quantization schedule and the synchronization cadence are chosen together rather than separately.

\subsection{Federated Alignment and Reinforcement Learning}
Our system currently supports supervised fine-tuning (Section~\ref{sec:finetune}). A natural next step is to support \emph{federated alignment}, since modern alignment pipelines rely heavily on reinforcement learning from human feedback (RLHF)~\cite{Ouyang:22:NIPS:RLHF}. RLHF does not fit easily into a federation: the reward model must be trained across institutions whose preference data is private and non-i.i.d., a setting in which centralized pipelines do not directly apply.
Preference-based objectives such as Direct Preference Optimization (DPO)~\cite{Rafailov:23:NIPS:DPO} may be better suited here, since they avoid explicit reward modeling and therefore reduce both the number of shared components and the volume of communication. Even so, preference data remains heterogeneous and institution-specific, and orchestrating such diverse preferences privately, without averaging away their diversity, is an open problem.

A further difficulty is \emph{distributed rollout generation}. RLHF repeatedly samples responses from the policy during training, and recent work identifies this generation step as the dominant systems bottleneck, which has motivated disaggregated execution~\cite{Xiao:24:preprint:FlexRLHF}. Performing rollouts across heterogeneous, WAN-connected sites, while keeping the sampled trajectories fresh enough for stable policy optimization, will require an execution model quite different from the co-located RLHF systems used today.

\subsection{Privacy, Security, and Trust}
Federated learning is often motivated by privacy. Nevertheless, the shared updates can still leak information about local data. Incorporating formal guarantees such as differential privacy (DP)~\cite{Wei:20:IEEETIFS, Xu:23:ACL} is therefore an important next step. The utility-privacy tradeoff at LLM scale is not yet well understood, since the noise level that a small model tolerates can substantially degrade a billion-parameter one.
Privacy is not the only concern. In an open federation whose participants are not all trusted, \emph{Byzantine robustness}, the ability to defend against clients that submit corrupted updates, becomes essential~\cite{Cao:21:NDSS:FLTrust}. Whether robust aggregation rules such as the coordinate-wise median or the trimmed mean retain their guarantees at LLM scale is an open question that our setting makes pressing.

\subsection{Incentive Design and Participation Economics}
A democratized training ecosystem works only if participants have a reason to contribute compute, data, and bandwidth. Existing schemes range from compute-centric rewards based on GPU hours or raw FLOPs~\cite{Pandey:22:NeurIPSW:FedToken} to Shapley-value methods~\cite{Ghorbani:19:ICML:DataShapley} that estimate each participant's marginal contribution to model quality.
Both overlook the \emph{temporal and staleness effects} that dominate asynchronous, geo-distributed training. A contribution that arrives during a critical convergence phase is worth far more than the same compute delivered after the loss has plateaued, yet neither metric distinguishes the two. Shapley scores are also unstable across aggregation strategies~\cite{Geimer:24:IJCNN:ShapleyVolatility}, and any deployable rule must account for strategic behavior by participants. Without incentives that reflect timing, data quality, and the possibility of gaming, a federation is vulnerable to free-riding and adverse selection. Designing incentive rules that are temporally aware, robust to manipulation, and inexpensive enough to enforce in real time is an open problem at the intersection of distributed systems and mechanism design.

\subsection{System-Algorithm Co-Design}
Building this system showed us that the algorithmic and systems choices cannot be made in isolation. The synchronization cadence, the compression strategy, and the fault-tolerance policy all interact with the optimizer, the learning-rate schedule, and the aggregation rule. Lengthening the synchronization interval, for example, saves communication but increases staleness, which may in turn call for a different outer optimizer, as DiLoCo~\cite{Douillard:23:DiLoCo} demonstrates with Nesterov momentum.
We therefore envision an \emph{auto-tuning} layer that monitors network conditions, compute heterogeneity, and convergence diagnostics, and adjusts the system parameters (compression ratio, synchronization frequency, timeout thresholds) together with the algorithmic hyperparameters (learning rate, local steps $H$, timeout $T$, aggregation weights). Such co-design would let the system adapt to the changing conditions of a geo-distributed deployment on its own, and would reduce the operational burden on participating institutions.

\subsection{Democratizing LLM Training at the Edge: \texorpdfstring{\\}{}A Call to Arms}
\label{sec:call-to-arms}

Democratizing LLM training is a multidisciplinary endeavor that cannot be accomplished by any single research community working in isolation. Realizing this vision requires rethinking the interplay between model training algorithms, distributed systems, communication networks, hardware architectures, and economic incentives. The challenges outlined above call for coordinated efforts across systems, machine learning, networking, security, and economics. Progress will depend not only on advances within each discipline, but also on principled approaches that bridge these disciplines to jointly optimize learning efficiency, system scalability, data privacy, robustness, and resource utilization.

Equally important is the development of shared open-source platforms, standardized benchmarks, and reproducible evaluation methodologies. Such infrastructure can lower barriers to entry, enable fair comparisons across competing approaches, and accelerate the translation of theoretical advances into practical solutions. By making tools, workloads, and experimental results broadly accessible, the research community can foster an ecosystem in which researchers, startups, smaller organizations, and resource-constrained communities can contribute to and benefit from advances in LLM training.

The stakes are high: Without concerted effort, the next generation of foundation models may remain the exclusive domain of a handful of well-resourced actors, further concentrating computational resources, technical expertise, and the capacity to shape AI development. Edge computing offers a promising opportunity to challenge this trajectory by harnessing distributed computational resources closer to where data are generated and applications are deployed. However, this opportunity will not be realized automatically. It demands fundamental advances in algorithms, systems, and infrastructure, together with sustained collaboration among academia, industry, open-source communities, and policymakers.

We believe the research community has both the technical capability and the collective responsibility to prevent the further concentration of LLM training capabilities. The time is ripe to move beyond isolated solutions toward a shared research agenda that makes LLM training more accessible, efficient, and inclusive. We therefore call on the community to treat democratizing LLM training at the edge not merely as a systems optimization problem, but as a broader scientific and societal challenge worthy of sustained, coordinated action.

%% file: bib/code.bib
@article{zhao2023fsdp,
  title={PyTorch FSDP: Experiences on Scaling Fully Sharded Data Parallel},
  author={Zhao, Yanli and Gu, Andrew and Varma, Rohan and Luo, Liang and Huang, Chien-Chin and Xu, Min and Wright, Less and Shojanazeri, Hamid and Ott, Myle and Shleifer, Sam and others},
  journal={arXiv preprint arXiv:2304.11277},
  year={2023}
}

@article{Beutel:20:preprint:flower,
    title={Flower: A Friendly Federated Learning Research Framework},
    author={Beutel, Daniel J and Topal, Taner and Mathur, Akhil and Qiu, Xinchi and Fernandez-Marques, Javier and Gao, Yan and Sani, Lorenzo and Kwing, Hei Li and Parcollet, Titouan and Gusmão, Pedro PB de and Lane, Nicholas D},
    journal={arXiv preprint arXiv:2007.14390},
    year={2020}
}

@article{Roth:23:IEEE:flare,
title = {{NVIDIA FLARE: Federated Learning from Simulation to Real-World}},
author = {Roth, Holger R. and Cheng, Yan and Wen, Yuhong and Yang, Isaac and Xu, Ziyue and Hsieh, Yuan-Ting and Kersten, Kristopher and Harouni, Ahmed and Zhao, Can and Lu, Kevin and Zhang, Zhihong and Li, Wenqi and Myronenko, Andriy and Yang, Dong and Yang, Sean and Rieke, Nicola and Quraini, Abood and Chen, Chester and Xu, Daguang and Ma, Nic and Dogra, Prerna and Flores, Mona and Feng, Andrew},
doi = {https://doi.org/10.48550/arXiv.2210.13291},
journal = {IEEE Data Eng. Bull., Vol. 46, No. 1},
month = mar,
year = {2023}
}

@article{Sani:24:preprint:photon,
    title={Photon: Federated LLM Pre-Training},
    author={Sani, Lorenzo and Iacob, Alex and Cao, Zeyu and Lee, Royson and Marino, Bill and Gao, Yan and Cai, Dongqi and Li, Zexi and Zhao, Wanru and Qiu, Xinchi and others},
    journal={arXiv preprint arXiv:2411.02908},
    year={2024}
}

@inproceedings{Borzunov:23:ACL:petals,
    title = {Petals: Collaborative Inference and Fine-tuning of Large Models},
    author = {Borzunov, Alexander and Baranchuk, Dmitry and Dettmers, Tim and Riabinin, Maksim and Belkada, Younes and Chumachenko, Artem and Samygin, Pavel and Raffel, Colin},
    booktitle = {Proceedings of the 61st Annual Meeting of the Association for Computational Linguistics (Volume 3: System Demonstrations)},
    pages = {558--568},
    year = {2023},
    url = {https://arxiv.org/abs/2209.01188}
}

@article{Reina:21:preprint:Openfl,
  title={Openfl: An open-source framework for federated learning},
  author={Reina, G Anthony and Gruzdev, Alexey and Foley, Patrick and Perepelkina, Olga and Sharma, Mansi and Davidyuk, Igor and Trushkin, Ilya and Radionov, Maksim and Mokrov, Aleksandr and Agapov, Dmitry and others},
  journal={arXiv preprint arXiv:2105.06413},
  year={2021}
}

@inproceedings{rajpurkar-etal-2016-squad,
    title = "{SQ}u{AD}: 100,000+ Questions for Machine Comprehension of Text",
    author = "Rajpurkar, Pranav  and
      Zhang, Jian  and
      Lopyrev, Konstantin  and
      Liang, Percy",
    editor = "Su, Jian  and
      Duh, Kevin  and
      Carreras, Xavier",
    booktitle = "Proceedings of the 2016 Conference on Empirical Methods in Natural Language Processing",
    month = nov,
    year = "2016",
    address = "Austin, Texas",
    publisher = "Association for Computational Linguistics",
    url = "https://aclanthology.org/D16-1264/",
    doi = "10.18653/v1/D16-1264",
    pages = "2383--2392"
}


%% file: bib/distrib_llm.bib
@inproceedings{Zhu:25:Mist,
    title={Mist: Efficient Distributed Training of Large Language Models via Memory-Parallelism Co-Optimization},
    author={Zhu, Zhanda and Giannoula, Christina and Andoorveedu, Muralidhar and Su, Qidong and Mangalam, Karttikeya and Zheng, Bojian and Pekhimenko, Gennady},
    booktitle={Proceedings of the Twentieth European Conference on Computer Systems},
    pages={1298--1316},
    year={2025}
}

@misc{Gugger:22:Accelerate,
    title =        {Accelerate: Training and inference at scale made simple, efficient and adaptable.},
    author =       {Gugger, Sylvain and Debut, Lysandre and Wolf, Thomas and Schmid, Philipp and Mueller, Zachary and Mangrulkar, Sourab},
    howpublished = {\url{https://github.com/huggingface/accelerate}},
    year =         {2022}
}

@inproceedings{Ye:24:openfedllm,
    title={Openfedllm: Training large language models on decentralized private data via federated learning},
    author={Ye, Rui and Wang, Wenhao and Chai, Jingyi and Li, Dihan and Li, Zexi and Xu, Yinda and Du, Yaxin and Wang, Yanfeng and Chen, Siheng},
    booktitle={Proceedings of the 30th ACM SIGKDD conference on knowledge discovery and data mining},
    pages={6137--6147},
    year={2024}
}

@inproceedings{Moritz:17:Ray,
  title={Ray: A distributed framework for emerging $\{$AI$\}$ applications},
  author={Moritz, Philipp and Nishihara, Robert and Wang, Stephanie and Tumanov, Alexey and Liaw, Richard and Liang, Eric and Elibol, Melih and Yang, Zongheng and Paul, William and Jordan, Michael I and others},
  booktitle={13th USENIX symposium on operating systems design and implementation (OSDI 18)},
  pages={561--577},
  year={2018}
}

@article{Karakus:21:SageMaker,
    title={Amazon sagemaker model parallelism: A general and flexible framework for large model training},
    author={Karakus, Can and Huilgol, Rahul and Wu, Fei and Subramanian, Anirudh and Daniel, Cade and Cavdar, Derya and Xu, Teng and Chen, Haohan and Rahnama, Arash and Quintela, Luis},
    journal={arXiv preprint arXiv:2111.05972},
    year={2021}
}

@book{Bhatia:23:VertexAI,
  title={The Definitive Guide to Google Vertex AI},
  author={Bhatia, Jasmeet and Chaudhary, Kartik},
  year={2023},
  publisher={Packt Publishing}
}

@article{Kuchaiev:19:NeMo,
    title={Nemo: a toolkit for building ai applications using neural modules},
    author={Kuchaiev, Oleksii and Li, Jason and Nguyen, Huyen and Hrinchuk, Oleksii and Leary, Ryan and Ginsburg, Boris and Kriman, Samuel and Beliaev, Stanislav and Lavrukhin, Vitaly and Cook, Jack and others},
    journal={arXiv preprint arXiv:1909.09577},
    year={2019}
}

@article{Jaghouar:24:PrimeIntellect,
  title={Intellect-1 technical report},
  author={Jaghouar, Sami and Ong, Jack Min and Basra, Manveer and Obeid, Fares and Straube, Jannik and Keiblinger, Michael and Bakouch, Elie and Atkins, Lucas and Panahi, Maziyar and Goddard, Charles and others},
  journal={arXiv preprint arXiv:2412.01152},
  year={2024}
}

@article{Cheng:25:preprint:Flock,
    title={Scaling Decentralized Learning with FLock},
    author={Cheng, Zehua and Sun, Rui and Sun, Jiahao and Guo, Yike},
    journal={arXiv preprint arXiv:2507.15349},
    year={2025}
}

@article{Singh:26:preprint:Knexa-FL,
    title={Learning to Collaborate: An Orchestrated-Decentralized Framework for Peer-to-Peer LLM Federation},
    author={Singh, Inderjeet and Vissol-Gaudin, Eleonore and Otung, Andikan and Sekiya, Motoyoshi},
    journal={arXiv preprint arXiv:2601.17133},
    year={2026}
}

@article{Qi:21:preprint:FedRLsurvey,
  title={Federated reinforcement learning: Techniques, applications, and open challenges},
  author={Qi, Jiaju and Zhou, Qihao and Lei, Lei and Zheng, Kan},
  journal={arXiv preprint arXiv:2108.11887},
  year={2021}
}

@article{Ouyang:22:NIPS:RLHF,
  title={Training language models to follow instructions with human feedback},
  author={Ouyang, Long and Wu, Jeffrey and Jiang, Xu and Almeida, Diogo and Wainwright, Carroll and Mishkin, Pamela and Zhang, Chong and Agarwal, Sandhini and Slama, Katarina and Ray, Alex and others},
  journal={Advances in neural information processing systems},
  volume={35},
  pages={27730--27744},
  year={2022}
}

@article{Rafailov:23:NIPS:DPO,
  title={Direct preference optimization: Your language model is secretly a reward model},
  author={Rafailov, Rafael and Sharma, Archit and Mitchell, Eric and Manning, Christopher D and Ermon, Stefano and Finn, Chelsea},
  journal={Advances in neural information processing systems},
  volume={36},
  pages={53728--53741},
  year={2023}
}

@inproceedings{Ahn:24:ACL:RLMathsurvey,
  title={Large language models for mathematical reasoning: Progresses and challenges},
  author={Ahn, Janice and Verma, Rishu and Lou, Renze and Liu, Di and Zhang, Rui and Yin, Wenpeng},
  booktitle={Proceedings of the 18th Conference of the European Chapter of the Association for Computational Linguistics: Student Research Workshop},
  pages={225--237},
  year={2024}
}

@article{Jiang:26:ACM:RLCodesurvey,
  title={A survey on large language models for code generation},
  author={Jiang, Juyong and Wang, Fan and Shen, Jiasi and Kim, Sungju and Kim, Sunghun},
  journal={ACM Transactions on Software Engineering and Methodology},
  volume={35},
  number={2},
  pages={1--72},
  year={2026},
  publisher={ACM New York, NY}
}

@article{Hou:25:ACM:MCP,
  title={Model context protocol (mcp): Landscape, security threats, and future research directions},
  author={Hou, Xinyi and Zhao, Yanjie and Wang, Shenao and Wang, Haoyu},
  journal={ACM Transactions on Software Engineering and Methodology},
  year={2025},
  publisher={ACM New York, NY}
}

@article{Gao:23:preprint:RAGsurvey,
  title={Retrieval-augmented generation for large language models: A survey},
  author={Gao, Yunfan and Xiong, Yun and Gao, Xinyu and Jia, Kangxiang and Pan, Jinliu and Bi, Yuxi and Dai, Yixin and Sun, Jiawei and Wang, Haofen and Wang, Haofen and others},
  journal={arXiv preprint arXiv:2312.10997},
  volume={2},
  number={1},
  pages={32},
  year={2023}
}


%% file: bib/fl.bib
@inproceedings{mcmahan2017fedavg,
  title={Communication-Efficient Learning of Deep Networks from Decentralized Data},
  author={McMahan, H. Brendan and Moore, Eider and Ramage, Daniel and Hampson, Seth and y Arcas, Blaise Aguera},
  booktitle={AISTATS},
  year={2017}
}

@inproceedings{li2020fedprox,
  title={Federated Optimization in Heterogeneous Networks},
  author={Li, Tian and Sahu, Anit Kumar and Zaheer, Manzil and Sanjabi, Maziar and Talwalkar, Ameet and Smith, Virginia},
  booktitle={MLSys},
  year={2020}
}

@inproceedings{Karimireddy:19:CoRR,
  title={Scaffold: Stochastic controlled averaging for federated learning},
  author={Karimireddy, Sai Praneeth and Kale, Satyen and Mohri, Mehryar and Reddi, Sashank and Stich, Sebastian and Suresh, Ananda Theertha},
  booktitle={International conference on machine learning},
  pages={5132--5143},
  year={2020},
  organization={PMLR}
}

@inproceedings{Reddi:20:ICLR,
    title={Adaptive Federated Optimization},
    author={Sashank J. Reddi and Zachary Charles and Manzil Zaheer and Zachary Garrett and Keith Rush and Jakub Kone{\v{c}}n{\'y} and Sanjiv Kumar and Hugh Brendan McMahan},
    booktitle={International Conference on Learning Representations},
    year={2021},
    url={https://openreview.net/forum?id=LkFG3lB13U5}
}

@inproceedings{Yang:21:ICLR,
    title={Achieving Linear Speedup with Partial Worker Participation in Non-{IID} Federated Learning},
    author={Haibo Yang and Minghong Fang and Jia Liu},
    booktitle={International Conference on Learning Representations},
    year={2021},
    url={https://openreview.net/forum?id=jDdzh5ul-d}
}

@inproceedings{Xu:23:ACL,
    title={Federated learning of gboard language models with differential privacy},
    author={Xu, Zheng and Zhang, Yanxiang and Andrew, Galen and Choquette, Christopher and Kairouz, Peter and Mcmahan, Brendan and Rosenstock, Jesse and Zhang, Yuanbo},
    booktitle={Proceedings of the 61st Annual Meeting of the Association for Computational Linguistics (Volume 5: Industry Track)},
    pages={629--639},
    year={2023}
}

@article{Wang:20:FedNova,
  title={Tackling the objective inconsistency problem in heterogeneous federated optimization},
  author={Wang, Jianyu and Liu, Qinghua and Liang, Hao and Joshi, Gauri and Poor, H Vincent},
  journal={Advances in neural information processing systems},
  volume={33},
  pages={7611--7623},
  year={2020}
}

@article{Douillard:23:DiLoCo,
    author       = {Arthur Douillard and
                    Qixuan Feng and
                    Andrei A. Rusu and
                    Rachita Chhaparia and
                    Yani Donchev and
                    Adhiguna Kuncoro and
                    Marc'Aurelio Ranzato and
                    Arthur Szlam and
                    Jiajun Shen},
    title        = {DiLoCo: Distributed Low-Communication Training of Language Models},
    journal      = {CoRR},
    volume       = {abs/2311.08105},
    year         = {2023},
    url          = {https://arxiv.org/abs/2311.08105},
    eprinttype   = {arXiv},
    eprint       = {2311.08105}
}

@inproceedings{Sutskever:13:ICML,
    title={On the importance of initialization and momentum in deep learning},
    author={Sutskever, Ilya and Martens, James and Dahl, George and Hinton, Geoffrey},
    booktitle={International conference on machine learning},
    pages={1139--1147},
    year={2013},
    organization={pmlr}
}

@inproceedings{Lalitha:18:NIPS,
    title={Fully decentralized federated learning},
    author={Lalitha, Anusha and Shekhar, Shubhanshu and Javidi, Tara and Koushanfar, Farinaz},
    booktitle={Third workshop on bayesian deep learning (NeurIPS)},
    volume={2},
    pages={1--9},
    year={2018}
}

@article{Savazzi:20:IEEEIoT,
    title={Federated learning with cooperating devices: A consensus approach for massive IoT networks},
    author={Savazzi, Stefano and Nicoli, Monica and Rampa, Vittorio},
    journal={IEEE Internet of Things Journal},
    volume={7},
    number={5},
    pages={4641--4654},
    year={2020},
    publisher={IEEE}
}

@inproceedings{Monschein:21:ISCC,
    title={Towards a peer-to-peer federated machine learning environment for continuous authentication},
    author={Monschein, David and P{\'e}rez, Jos{\'e} Antonio Peregrina and Piotrowski, Tim and Nochta, Zolt{\'a}n and Waldhorst, Oliver P and Zirpins, Christian},
    booktitle={2021 IEEE Symposium on computers and communications (ISCC)},
    pages={1--6},
    year={2021},
    organization={IEEE}
}

@article{Chellapandi:23:IEEECSL,
    title={On the convergence of decentralized federated learning under imperfect information sharing},
    author={Chellapandi, Vishnu Pandi and Upadhyay, Antesh and Hashemi, Abolfazl and {\.Z}ak, Stanislaw H},
    journal={IEEE Control Systems Letters},
    volume={7},
    pages={2982--2987},
    year={2023},
    publisher={IEEE}
}

@article{Lim:20:FLsurvey,
    title={Federated learning in mobile edge networks: A comprehensive survey},
    author={Lim, Wei Yang Bryan and Luong, Nguyen Cong and Hoang, Dinh Thai and Jiao, Yutao and Liang, Ying-Chang and Yang, Qiang and Niyato, Dusit and Miao, Chunyan},
    journal={IEEE communications surveys \& tutorials},
    volume={22},
    number={3},
    pages={2031--2063},
    year={2020},
    publisher={IEEE}
}

@article{Yuan:24:DFLsurvey,
    title={Decentralized federated learning: A survey and perspective},
    author={Yuan, Liangqi and Wang, Ziran and Sun, Lichao and Yu, Philip S and Brinton, Christopher G},
    journal={IEEE Internet of Things Journal},
    volume={11},
    number={21},
    pages={34617--34638},
    year={2024},
    publisher={IEEE}
}

@article{Wei:20:IEEETIFS,
  title={Federated learning with differential privacy: Algorithms and performance analysis},
  author={Wei, Kang and Li, Jun and Ding, Ming and Ma, Chuan and Yang, Howard H and Farokhi, Farhad and Jin, Shi and Quek, Tony QS and Poor, H Vincent},
  journal={IEEE transactions on information forensics and security},
  volume={15},
  pages={3454--3469},
  year={2020},
  publisher={IEEE}
}


%% file: bib/foundation.bib
@misc{Touvron:23:LLaMA,
    title={LLaMA: Open and Efficient Foundation Language Models}, 
    author={Hugo Touvron and Thibaut Lavril and Gautier Izacard and Xavier Martinet and Marie-Anne Lachaux and Timothée Lacroix and Baptiste Rozière and Naman Goyal and Eric Hambro and Faisal Azhar and Aurelien Rodriguez and Armand Joulin and Edouard Grave and Guillaume Lample},
    year={2023},
    eprint={2302.13971},
    archivePrefix={arXiv},
    primaryClass={cs.CL},
    url={https://arxiv.org/abs/2302.13971}, 
}

@article{Radford:19:OpenAI,
    title={Language models are unsupervised multitask learners},
    author={Radford, Alec and Wu, Jeffrey and Child, Rewon and Luan, David and Amodei, Dario and Sutskever, Ilya and others},
    journal={OpenAI blog},
    volume={1},
    number={8},
    pages={9},
    year={2019}
}

@inproceedings{Devlin:19:BERT,
    title={Bert: Pre-training of deep bidirectional transformers for language understanding},
    author={Devlin, Jacob and Chang, Ming-Wei and Lee, Kenton and Toutanova, Kristina},
    booktitle={Proceedings of the 2019 conference of the North American chapter of the association for computational linguistics: human language technologies, volume 1 (long and short papers)},
    pages={4171--4186},
    year={2019}
}

@misc{OpenAI:24:GPT4,
    title={GPT-4 Technical Report}, 
    author={OpenAI and Josh Achiam and others},
    year={2024},
    eprint={2303.08774},
    archivePrefix={arXiv},
    primaryClass={cs.CL},
    url={https://arxiv.org/abs/2303.08774}, 
}

@inproceedings{Radford:21:CLIP,
  title={Learning transferable visual models from natural language supervision},
  author =       {Alec Radford and Jong Wook Kim and Chris Hallacy and Aditya Ramesh and Gabriel Goh and Sandhini Agarwal and Girish Sastry and Amanda Askell and Pamela Mishkin and Jack Clark and Gretchen Krueger and Ilya Sutskever},
  booktitle={International conference on machine learning},
  pages={8748--8763},
  year={2021},
  organization={PMLR}
}

@misc{Ramesh:22:DALLE2,
    title={Hierarchical Text-Conditional Image Generation with CLIP Latents}, 
    author={Aditya Ramesh and Prafulla Dhariwal and Alex Nichol and Casey Chu and Mark Chen},
    year={2022},
    eprint={2204.06125},
    archivePrefix={arXiv},
    primaryClass={cs.CV},
    url={https://arxiv.org/abs/2204.06125}, 
}

@article{Vaswani:17:NIPS,
    title={Attention is all you need},
    author={Vaswani, Ashish and Shazeer, Noam and Parmar, Niki and Uszkoreit, Jakob and Jones, Llion and Gomez, Aidan N and Kaiser, {\L}ukasz and Polosukhin, Illia},
    journal={Advances in neural information processing systems},
    volume={30},
    year={2017}
}

@inproceedings{Groeneveld:24:ACL,
    title={OLMo: Accelerating the science of language models},
    author={Groeneveld, Dirk and Beltagy, Iz and Walsh, Evan and Bhagia, Akshita and Kinney, Rodney and Tafjord, Oyvind and Jha, Ananya and Ivison, Hamish and Magnusson, Ian and Wang, Yizhong and others},
    booktitle={Proceedings of the 62nd Annual Meeting of the Association for Computational Linguistics (Volume 1: Long Papers)},
    pages={15789--15809},
    year={2024}
}

@article{Brown:20:NIPS,
    title={Language models are few-shot learners},
    author={Brown, Tom and Mann, Benjamin and Ryder, Nick and Subbiah, Melanie and Kaplan, Jared D and Dhariwal, Prafulla and Neelakantan, Arvind and Shyam, Pranav and Sastry, Girish and Askell, Amanda and others},
    journal={Advances in neural information processing systems},
    volume={33},
    pages={1877--1901},
    year={2020}
}

@inproceedings{Shazeer:17:ICLR,
    title={Outrageously Large Neural Networks: The Sparsely-Gated Mixture-of-Experts Layer},
    author={Noam Shazeer and *Azalia Mirhoseini and *Krzysztof Maziarz and Andy Davis and Quoc Le and Geoffrey Hinton and Jeff Dean},
    booktitle={International Conference on Learning Representations},
    year={2017},
    url={https://openreview.net/forum?id=B1ckMDqlg}
}

@article{Liu:24:preprint,
    title={Deepseek-v3 technical report},
    author={Liu, Aixin and Feng, Bei and Xue, Bing and Wang, Bingxuan and Wu, Bochao and Lu, Chengda and Zhao, Chenggang and Deng, Chengqi and Zhang, Chenyu and Ruan, Chong and others},
    journal={arXiv preprint arXiv:2412.19437},
    year={2024}
}

@inproceedings{Kandpal:25:ICML,
    title={Position: The Most Expensive Part of an {LLM} *should* be its Training Data},
    author={Nikhil Kandpal and Colin Raffel},
    booktitle={Forty-second International Conference on Machine Learning Position Paper Track},
    year={2025},
    url={https://openreview.net/forum?id=L6RpQ1h4Nx}
}


%% file: bib/future_directions.bib
@inproceedings{Thorpe:23:NSDI:Bamboo,
  title     = {Bamboo: Making Preemptible Instances Resilient for Affordable Training of Large {DNNs}},
  author    = {Thorpe, John and Zhao, Pengzhan and Eyolfson, Jonathan and Qiao, Yifan and Jia, Zhihao and Zhang, Minjia and Netravali, Ravi and Xu, Guoqing Harry},
  booktitle = {20th USENIX Symposium on Networked Systems Design and Implementation},
  year      = {2023}
}

@inproceedings{Jang:23:SOSP:Oobleck,
  title     = {Oobleck: Resilient Distributed Training of Large Models Using Pipeline Templates},
  author    = {Jang, Insu and Yang, Zhenning and Zhang, Zhen and Jin, Xin and Chowdhury, Mosharaf},
  booktitle = {ACM SIGOPS 29th Symposium on Operating Systems Principles},
  year      = {2023}
}

@inproceedings{Wang:23:SOSP:Gemini,
  title     = {{Gemini}: Fast Failure Recovery in Distributed Training with In-Memory Checkpoints},
  author    = {Wang, Zhuang and Jia, Zhen and Zheng, Shuai and Zhang, Zhen and Fu, Xinwei and Ng, T.\ S.\ Eugene and You, Yang},
  booktitle = {ACM SIGOPS 29th Symposium on Operating Systems Principles},
  year      = {2023}
}

@inproceedings{Pandey:22:NeurIPSW:FedToken,
  title     = {{FedToken}: Tokenized Incentives for Data Contribution in Federated Learning},
  author    = {Pandey, Shashi Raj and Nguyen, Lam Duc and Popovski, Petar},
  booktitle = {Workshop on Federated Learning: Recent Advances and New Challenges (FL-NeurIPS)},
  year      = {2022}
}

@inproceedings{Ghorbani:19:ICML:DataShapley,
  title     = {Data {Shapley}: Equitable Valuation of Data for Machine Learning},
  author    = {Ghorbani, Amirata and Zou, James},
  booktitle = {International Conference on Machine Learning},
  pages     = {2242--2251},
  year      = {2019},
  organization = {PMLR}
}

@inproceedings{Geimer:24:IJCNN:ShapleyVolatility,
  title     = {On the Volatility of {Shapley}-Based Contribution Metrics in Federated Learning},
  author    = {Geimer, Arno and Fiz, Beltran and State, Radu},
  booktitle = {International Joint Conference on Neural Networks},
  year      = {2025}
}

@inproceedings{Cao:21:NDSS:FLTrust,
  title={{FLTrust}: {Byzantine}-Robust Federated Learning via Trust Bootstrapping},
  author={Cao, Xiaoyu and Fang, Minghong and Liu, Jia and Gong, Neil Zhenqiang},
  booktitle={Network and Distributed System Security Symposium},
  year={2021}
}

@article{Xiao:24:preprint:FlexRLHF,
  title={An Adaptive Placement and Parallelism Framework for Accelerating {RLHF} Training},
  author={Xiao, Youshao and Zhou, Zhenglei and Mao, Fagui and Wu, Weichang and Zhao, Shangchun and Ju, Lin and Liang, Lei and Zhang, Xiaolu and Zhou, Jun},
  journal={arXiv preprint arXiv:2312.11819},
  year={2024}
}

@inproceedings{Ryabinin:23:ICML:SWARM,
  title={{{SWARM} Parallelism: Training Large Models Can Be Surprisingly Communication-Efficient}},
  author={Ryabinin, Max and Dettmers, Tim and Diskin, Michael and Borzunov, Alexander},
  booktitle={International Conference on Machine Learning},
  pages={29416--29440},
  year={2023},
  organization={PMLR}
}

@article{RamezaniKebrya:21:JMLR:NUQSGD,
  title={{NUQSGD}: Provably Communication-Efficient Data-Parallel {SGD} via Nonuniform Quantization},
  author={Ramezani-Kebrya, Ali and Faghri, Fartash and Markov, Ilya and Aksenov, Vitalii and Alistarh, Dan and Roy, Daniel M.},
  journal={Journal of Machine Learning Research},
  volume={22},
  number={114},
  pages={1--43},
  year={2021}
}

@inproceedings{Liu:23:NeurIPS:LLaVA,
  title={Visual Instruction Tuning},
  author={Liu, Haotian and Li, Chunyuan and Wu, Qingyang and Lee, Yong Jae},
  booktitle={Advances in Neural Information Processing Systems},
  year={2023}
}


%% file: bib/general.bib
@article{Kaplan:20:preprint,
    title={Scaling laws for neural language models},
    author={Kaplan, Jared and McCandlish, Sam and Henighan, Tom and Brown, Tom B and Chess, Benjamin and Child, Rewon and Gray, Scott and Radford, Alec and Wu, Jeffrey and Amodei, Dario},
    journal={arXiv preprint arXiv:2001.08361},
    year={2020}
}

@article{Hoffmann:22:NIPS,
    title={An empirical analysis of compute-optimal large language model training},
    author={Hoffmann, Jordan and Borgeaud, Sebastian and Mensch, Arthur and Buchatskaya, Elena and Cai, Trevor and Rutherford, Eliza and de Las Casas, Diego and Hendricks, Lisa Anne and Welbl, Johannes and Clark, Aidan and others},
    journal={Advances in neural information processing systems},
    volume={35},
    pages={30016--30030},
    year={2022}
}

@inproceedings{
hu2022lora,
title={Lo{RA}: Low-Rank Adaptation of Large Language Models},
author={Edward J Hu and yelong shen and Phillip Wallis and Zeyuan Allen-Zhu and Yuanzhi Li and Shean Wang and Lu Wang and Weizhu Chen},
booktitle={International Conference on Learning Representations},
year={2022},
url={https://openreview.net/forum?id=nZeVKeeFYf9}
}

@article{Dettmers:23:NIPS:Qlora,
    title={Qlora: Efficient finetuning of quantized llms},
    author={Dettmers, Tim and Pagnoni, Artidoro and Holtzman, Ari and Zettlemoyer, Luke},
    journal={Advances in neural information processing systems},
    volume={36},
    pages={10088--10115},
    year={2023}
}

@inproceedings{Li:21:ACL:prefixtune,
    title={Prefix-tuning: Optimizing continuous prompts for generation},
    author={Li, Xiang Lisa and Liang, Percy},
    booktitle={Proceedings of the 59th Annual Meeting of the Association for Computational Linguistics and the 11th International Joint Conference on Natural Language Processing (Volume 1: Long Papers)},
    pages={4582--4597},
    year={2021}
}

@inproceedings{Ben:21:ACL:BitFit,
    title = "{B}it{F}it: Simple Parameter-efficient Fine-tuning for Transformer-based Masked Language-models",
    author = "Ben Zaken, Elad  and
      Goldberg, Yoav  and
      Ravfogel, Shauli",
    editor = "Muresan, Smaranda  and
      Nakov, Preslav  and
      Villavicencio, Aline",
    booktitle = "Proceedings of the 60th Annual Meeting of the {A}ssociation for {C}omputational {L}inguistics (Volume 2: Short Papers)",
    month = may,
    year = "2022",
    address = "Dublin, Ireland",
    publisher = "Association for Computational Linguistics",
    url = "https://aclanthology.org/2022.acl-short.1/",
    doi = "10.18653/v1/2022.acl-short.1",
    pages = "1--9"
}

@inproceedings{Pfeiffer:21:ACL:adapterfusion,
    title={Adapterfusion: Non-destructive task composition for transfer learning},
    author={Pfeiffer, Jonas and Kamath, Aishwarya and R{\"u}ckl{\'e}, Andreas and Cho, Kyunghyun and Gurevych, Iryna},
    booktitle={Proceedings of the 16th conference of the European chapter of the association for computational linguistics: main volume},
    pages={487--503},
    year={2021}
}

@inproceedings{Sun:24:ICLR,
    title={Improving Lo{RA} in Privacy-preserving Federated Learning},
    author={Youbang Sun and Zitao Li and Yaliang Li and Bolin Ding},
    booktitle={The Twelfth International Conference on Learning Representations},
    year={2024},
    url={https://openreview.net/forum?id=NLPzL6HWNl}
}

@inproceedings{Wang:24:NIPS,
    title={{FL}o{RA}: Federated Fine-Tuning Large Language Models with Heterogeneous Low-Rank Adaptations},
    author={Ziyao Wang and Zheyu Shen and Yexiao He and Guoheng Sun and Hongyi Wang and Lingjuan Lyu and Ang Li},
    booktitle={The Thirty-eighth Annual Conference on Neural Information Processing Systems},
    year={2024},
    url={https://openreview.net/forum?id=TcCorXxNJQ}
}

@article{
wu2026asurvey,
title={A Survey on Federated Fine-Tuning of Large Language Models},
author={Yebo Wu and Chunlin Tian and Jingguang Li and He Sun and KaHou Tam and Zhanting Zhou and Haicheng Liao and Jing Xiong and Zhijiang Guo and Li Li and Cheng-zhong Xu},
journal={Transactions on Machine Learning Research},
issn={2835-8856},
year={2026},
url={https://openreview.net/forum?id=rnCqbuIWnn},
note={Survey Certification}
}

@article{Bai:24:NIPS:FlexLora,
    title={Federated fine-tuning of large language models under heterogeneous tasks and client resources},
    author={Bai, Jiamu and Chen, Daoyuan and Qian, Bingchen and Yao, Liuyi and Li, Yaliang},
    journal={Advances in Neural Information Processing Systems},
    volume={37},
    pages={14457--14483},
    year={2024}
}

@article{Yan:25:IEEEComm,
    title={Federated fine-tuning of llms: Framework comparison and research directions},
    author={Yan, Na and Su, Yang and Deng, Yansha and Schober, Robert},
    journal={IEEE Communications Magazine},
    volume={63},
    number={10},
    pages={52--58},
    year={2025},
    publisher={IEEE}
}

@inproceedings{Zhang:24:ICASSP:FedIT,
  title={Towards building the federatedgpt: Federated instruction tuning},
  author={Zhang, Jianyi and Vahidian, Saeed and Kuo, Martin and Li, Chunyuan and Zhang, Ruiyi and Yu, Tong and Wang, Guoyin and Chen, Yiran},
  booktitle={ICASSP 2024-2024 IEEE international conference on acoustics, speech and signal processing (ICASSP)},
  pages={6915--6919},
  year={2024},
  organization={IEEE}
}

@inproceedings{Raje:25:NIPS:RAVAN,
    title={Ravan: Multi-Head Low-Rank Adaptation for Federated Fine-Tuning},
    author={Arian Raje and Baris Askin and Divyansh Jhunjhunwala and Gauri Joshi},
    booktitle={The Thirty-ninth Annual Conference on Neural Information Processing Systems},
    year={2025},
    url={https://openreview.net/forum?id=gyn4n8oC9B}
}

@inproceedings{Li:25:ICLR:ZeroOrder,
    title={Achieving Dimension-Free Communication in Federated Learning via Zeroth-Order Optimization},
    author={Zhe Li and Bicheng Ying and Zidong Liu and  Chaosheng Dong and Haibo Yang},
    booktitle={The Thirteenth International Conference on Learning Representations},
    year={2025},
    url={https://openreview.net/forum?id=omrLHFzC37}
}

@inproceedings{Villalobos:24:ICML,
    author = {Villalobos, Pablo and Ho, Anson and Sevilla, Jaime and Besiroglu, Tamay and Heim, Lennart and Hobbhahn, Marius},
    title = {Position: will we run out of data? limits of LLM scaling based on human-generated data},
    year = {2024},
    publisher = {JMLR.org},
    booktitle = {Proceedings of the 41st International Conference on Machine Learning},
    articleno = {2024},
    numpages = {22},
    location = {Vienna, Austria},
    series = {ICML'24}
}

@misc{OSC:25:web,
    author = {{Ohio Supercomputer Center}},
    title  = {Cluster Computing},
    url    = {https://www.osc.edu/resources/technical_support/supercomputers},
    year   = {2025}
}

@inproceedings{Strati:25:SOSP,
    title={{Sailor: Automating Distributed Training over Dynamic, Heterogeneous, and Geo-distributed Clusters}},
    author={Strati, Foteini and Zhang, Zhendong and Manos, George and P{\'e}riz, Ixeia S{\'a}nchez and Hu, Qinghao and Chen, Tiancheng and Buzcu, Berk and Han, Song and Delgado, Pamela and Klimovic, Ana},
    booktitle={Proceedings of the ACM SIGOPS 31st Symposium on Operating Systems Principles},
    pages={204--220},
    year={2025}
}


%% file: bib/prelim.bib
@inproceedings{dean2012distbelief,
  title={Large Scale Distributed Deep Networks},
  author={Dean, Jeff and Corrado, Greg and Monga, Rajat and Chen, Kai and Devin, Matthieu and Le, Quoc and Mao, Mark and Ranzato, Marc'Aurelio and Senior, Andrew and Tucker, Paul and others},
  booktitle={NeurIPS},
  year={2012}
}

@inproceedings{li2014ps,
  title={Scaling Distributed Machine Learning with the Parameter Server},
  author={Li, Mu and Andersen, David G and Park, Jun Woo and Smola, Alex and Ahmed, Amr and Josifovski, Vanja and Long, James and Shekita, Eugene J and Su, Bor-Yiing},
  booktitle={OSDI},
  year={2014}
}

@techreport{ho2013ssp,
  title={More Effective Distributed ML via a Stale Synchronous Parallel Parameter Server},
  author={Ho, Qirong and Cipar, James and Cui, Henggang and Lee, Jin Kyu and Kim, Joon Kyu and Gibbons, Phillip B and Gibson, Garth and Ganger, Gregory R and Xing, Eric P},
  institution={CMU},
  year={2013}
}

@inproceedings{bonawitz2019system,
  title={Towards Federated Learning at Scale: System Design},
  author={Bonawitz, Keith and Eichner, Hubert and Grieskamp, Wolfgang and Huba, Dzmitry and Ingerman, Alex and Ivanov, Vladimir and Kiddon, Chlo{\'e} and Konecn{\'y}, Jakub and Mazzocchi, Stefano and McMahan, H. Brendan and others},
  booktitle={MLSys},
  year={2019}
}

@article{xie2019asyncfl,
  title={Asynchronous Federated Optimization},
  author={Xie, Cong and Koyejo, Sanmi and Gupta, Indranil},
  journal={arXiv preprint arXiv:1903.03934},
  year={2019}
}

@inproceedings{nguyen2022fedbuff,
  title={Federated Learning with Buffered Asynchronous Aggregation},
  author={Nguyen, John and Malik, Kshitiz and Zhan, Hongyuan and Yousefpour, Ashkan and Rabbat, Michael and Malek, Mani and Huba, Dzmitry},
  booktitle={AISTATS},
  year={2022}
}

@article{stich2018localsgd,
  title={Local SGD Converges Fast and Communicates Little},
  author={Stich, Sebastian U.},
  journal={arXiv preprint arXiv:1805.09767},
  year={2018}
}

@article{rajbhandari2020zero,
  title={ZeRO: Memory Optimizations Toward Training Trillion Parameter Models},
  author={Rajbhandari, Samyam and Rasley, Jeff and Ruwase, Olatunji and He, Yuxiong},
  journal={arXiv preprint arXiv:1910.02054},
  year={2020}
}

@article{shoeybi2020megatron,
  title={Megatron-LM: Training Multi-Billion Parameter Language Models Using Model Parallelism},
  author={Shoeybi, Mohammad and Patwary, Mostofa and Puri, Raul and LeGresley, Patrick and Casper, Jared and Catanzaro, Bryan},
  journal={arXiv preprint arXiv:1909.08053},
  year={2020}
}

@inproceedings{zheng2022alpa,
  title={Alpa: Automating Inter- and Intra-Operator Parallelism for Distributed Deep Learning},
  author={Zheng, Lianmin and others},
  booktitle={OSDI},
  year={2022}
}

@article{konevcny2016federated,
  title={Federated Optimization: Distributed Machine Learning for On-Device Intelligence},
  author={Kone{\v{c}}n{\`y}, Jakub and McMahan, H. Brendan and Ramage, Daniel and Richt{\'a}rik, Peter},
  journal={arXiv preprint arXiv:1610.02527},
  year={2016}
}

@inproceedings{recht2011hogwild,
  title={Hogwild!: A Lock-Free Approach to Parallelizing Stochastic Gradient Descent},
  author={Recht, Benjamin and R{\'e}, Christopher and Wright, Stephen and Niu, Feng},
  booktitle={Advances in Neural Information Processing Systems},
  year={2011}
}

@article{koloskova2022sharper,
  title={Sharper Convergence Guarantees for Asynchronous SGD for Distributed and Federated Learning},
  author={Koloskova, Anastasia and Stich, Sebastian U. and Jaggi, Martin},
  journal={Advances in Neural Information Processing Systems},
  year={2022}
}

@article{chazan1969chaotic,
  title={Chaotic Relaxation},
  author={Chazan, D. and Miranker, W.},
  journal={Linear Algebra and its Applications},
  volume={2},
  number={2},
  pages={199--222},
  year={1969}
}

@article{tsitsiklis1986distributed,
  title={Distributed Asynchronous Deterministic and Stochastic Gradient Optimization Algorithms},
  author={Tsitsiklis, John N. and Bertsekas, Dimitri P. and Athans, Michael},
  journal={IEEE Transactions on Automatic Control},
  volume={31},
  number={9},
  pages={803--812},
  year={1986}
}

@book{bertsekas1989parallel,
  title={Parallel and Distributed Computation: Numerical Methods},
  author={Bertsekas, Dimitri P. and Tsitsiklis, John N.},
  publisher={Prentice Hall},
  year={1989}
}

@inproceedings{jiang2020byteps,
  title={A Unified Architecture for Accelerating Distributed DNN Training in Heterogeneous GPU/CPU Clusters},
  author={Jiang, Yibo and et al.},
  booktitle={OSDI},
  year={2020}
}

@inproceedings{chai2020tifl,
  title={TiFL: A Tier-based Federated Learning System},
  author={Chai, Zheng and Ali, Ahsan and Zawad, Syed and Truex, Stacey and Anwar, Ali and Baracaldo, Nathalie and Zhou, Yi and Ludwig, Heiko and Yan, Feng and Cheng, Yue},
  booktitle={HPDC},
  year={2020}
}

@inproceedings{lai2021oort,
  title={Oort: Efficient Federated Learning via Guided Participant Selection},
  author={Lai, Fan and Zhu, Xiangfeng and Madhyastha, Harsha V. and Chowdhury, Mosharaf},
  booktitle={OSDI},
  year={2021}
}

@inproceedings{lai2022fedscale,
  title={FedScale: Benchmarking Model and System Performance of Federated Learning at Scale},
  author={Lai, Fan and Dai, Yinwei and Singapuram, Sanjay and Liu, Jiachen and Zhu, Xiangfeng and Madhyastha, Harsha and Chowdhury, Mosharaf},
  booktitle={ICML},
  year={2022}
}

@inproceedings{
nourzad2026one,
title={One Model, Many Goals: Meta-Learning Preference-Conditioned Alignment for Lifelong {LLM} Agents},
author={Fatemeh Nourzad and Daouda Sow and Yingbin Liang and Ming Shi and Ming Zhang and Yunxuan Li and Eylem Ekici and Ness Shroff},
booktitle={ICLR 2026 Workshop on Lifelong Agents: Learning, Aligning, Evolving},
year={2026},
url={https://openreview.net/forum?id=2FVShQSae2}
}

@inproceedings{nourzadbarriers,
  title     = {Barriers to Pareto Steerability in Preference-Conditioned LLM Alignment},
  author    = {Nourzad, Fatemeh and Sow, Daouda and Liang, Yingbin and Shi, Ming and Zhang, Ming and Li, Yunxuan and Ekici, Eylem and Shroff, Ness},
  booktitle = {ICLR 2026 Workshop on I Can't Believe It's Not Better},
  year      = {2026},
  url       = {https://openreview.net/forum?id=GVlh3flsqF}
}

@inproceedings{wang2020fednova,
  title={Tackling the Objective Inconsistency Problem in Heterogeneous Federated Optimization},
  author={Wang, Jianyu and Liu, Qinghua and Liang, Hao and Joshi, Gauri and Poor, H. Vincent},
  booktitle={NeurIPS},
  year={2020}
}

@inproceedings{karimireddy2021mime,
  title={MIME: Mimicking Centralized Stochastic Algorithms in Federated Learning},
  author={Karimireddy, Sai Praneeth and Reddi, Sashank J. and Stich, Sebastian U. and Jaggi, Martin and Kale, Satyen and Kumar, Sanjiv},
  booktitle={ICML},
  year={2021}
}

@misc{nourzad2026firm,
      title={FIRM: Federated In-client Regularized Multi-objective Alignment for Large Language Models}, 
      author={Fatemeh Nourzad and Amirhossein Roknilamouki and Eylem Ekici and Jia Liu and Ness Shroff},
      year={2026},
      eprint={2511.16992},
      archivePrefix={arXiv},
      primaryClass={cs.LG},
      url={https://arxiv.org/abs/2511.16992}, 
}


%% file: bib/split_learn.bib
@inproceedings{
    Nair:25:ICML,
    title={{FSL}-{SAGE}: Accelerating Federated Split Learning via Smashed Activation Gradient Estimation},
    author={Srijith Nair and Michael Lin and Peizhong Ju and Amirreza Talebi and Elizabeth Serena Bentley and Jia Liu},
    booktitle={Forty-second International Conference on Machine Learning},
    year={2025},
    url={https://openreview.net/forum?id=HnwcrtoDd4}
}

@inproceedings{Thapa:22:AAAI,
    title={Splitfed: When federated learning meets split learning},
    author={Thapa, Chandra and Arachchige, Pathum Chamikara Mahawaga and Camtepe, Seyit and Sun, Lichao},
    booktitle={Proceedings of the AAAI Conference on Artificial Intelligence},
    volume={36},
    pages={8485--8493},
    year={2022}
}

@article{Vepakomma:18:preprint,
    title={Split learning for health: Distributed deep learning without sharing raw patient data},
    author={Vepakomma, Praneeth and Gupta, Otkrist and Swedish, Tristan and Raskar, Ramesh},
    journal={arXiv preprint arXiv:1812.00564},
    year={2018}
}

@inproceedings{Harlap:19:sosp:PipeDream,
author = {Narayanan, Deepak and Harlap, Aaron and Phanishayee, Amar and Seshadri, Vivek and Devanur, Nikhil R. and Ganger, Gregory R. and Gibbons, Phillip B. and Zaharia, Matei},
title = {PipeDream: generalized pipeline parallelism for DNN training},
year = {2019},
isbn = {9781450368735},
publisher = {Association for Computing Machinery},
address = {New York, NY, USA},
url = {https://doi.org/10.1145/3341301.3359646},
doi = {10.1145/3341301.3359646},
booktitle = {Proceedings of the 27th ACM Symposium on Operating Systems Principles},
pages = {1-15},
numpages = {15},
location = {Huntsville, Ontario, Canada},
series = {SOSP '19}
}

@article{Huang:19:NIPS:GPipe,
    title={Gpipe: Efficient training of giant neural networks using pipeline parallelism},
    author={Huang, Yanping and Cheng, Youlong and Bapna, Ankur and Firat, Orhan and Chen, Dehao and Chen, Mia and Lee, HyoukJoong and Ngiam, Jiquan and Le, Quoc V and Wu, Yonghui and others},
    journal={Advances in neural information processing systems},
    volume={32},
    year={2019}
}
